\documentclass[fleqn,usenatbib]{mnras}

\usepackage{newtxtext,newtxmath}

\usepackage[T1]{fontenc}

\usepackage[normalem]{ulem}  
\newcommand\redout{\bgroup\markoverwith
{\textcolor{red}{\rule[0.5ex]{2pt}{0.8pt}}}\ULon}

\usepackage{tikz,xcolor,hyperref}

\definecolor{lime}{HTML}{A6CE39}
\DeclareRobustCommand{\orcidicon}{
	\begin{tikzpicture}
	\draw[lime, fill=lime] (0,0) 
	circle [radius=0.16] 
	node[white] {{\fontfamily{qag}\selectfont \tiny ID}};
	\draw[white, fill=white] (-0.0625,0.095) 
	circle [radius=0.007];
	\end{tikzpicture}
	\hspace{-2mm}
}

\foreach \x in {A, ..., Z}{\expandafter\xdef\csname orcid\x\endcsname{\noexpand\href{https://orcid.org/\csname orcidauthor\x\endcsname}
			{\noexpand\orcidicon}}
}

\usepackage{graphicx}	
\usepackage{amsmath}	
\usepackage{pdflscape} 
\usepackage{slantsc}

\title[The power of binaries on stripped-envelope supernovae across metallicity]{The power of binaries on stripped-envelope supernovae across metallicity: uniform progenitor parameter space and persistently low ejecta masses, but subtype diversity}

\author[D. Souropanis et al.]{
D. Souropanis$^{1\orcidA{}}$\thanks{E-mail: dsouropanis@ia.forth.gr},
E. Zapartas$^{1}$, T. Pessi$^{2,3}$, M. Briel$^{4,5}$,  M. Renzo$^{6}$, C. P. Gutiérrez$^{7,8}$, J. J. Andrews$^{9,10}$, \newauthor   S. Gossage$^{11,12}$, 
M. U. Kruckow$^{4,5}$, C. Liotine$^{11,12}$,  P. M. Srivastava$^{11,12,13}$,  and E. Teng$^{11,12,14}$
\\
\\
\\
$^{1}$ Institute of Astrophysics, Foundation for Research and Technology-Hellas, GR-71110 Heraklion, Greece \\
$^{2}$European Southern Observatory, Alonso de Córdova 3107, Vitacura, Casilla 19001, Santiago, Chile\\
$^{3}$ Instituto de Estudios Astrofísicos, Facultad de Ingeniería y Ciencias, Universidad Diego Portales, Av. Ejército Libertador 441,
Santiago, Chile\\
$^{4}$D\'epartement d’Astronomie, Universit\'e de Gen\`eve, Versoix, Switzerland\\
$^{5}$Gravitational Wave Science Center (GWSC), Universit\'e de Gen\`eve, Geneva, Switzerland\\
$^{6}$ University of Arizona, Department of Astronomy \& Steward Observatory, 933 N. Cherry Ave., Tucson, AZ 85721, USA\\
$^{7}$ Institut d’Estudis Espacials de Catalunya (IEEC), 08860 Castelldefels (Barcelona), Spain\\
$^{8}$ Institute of Space Sciences (ICE, CSIC), Campus UAB, Carrer de Can Magrans, s/n, E-08193 Barcelona, Spain\\
$^{9}$Department of Physics, University of Florida, Gainesville, FL, USA\\
$^{10}$Institute for Fundamental Theory, University of Florida, Gainesville, FL, USA\\
$^{11}$Center for Interdisciplinary Exploration and Research in Astrophysics (CIERA), Northwestern University, Evanston, IL, USA\\
$^{12}$NSF\textendash Simons AI Institute for the Sky (SkAI), Chicago, IL, USA\\
$^{13}$Department of Electrical and Computer Engineering, Northwestern University, Evanston, IL, USA\\
$^{14}$Department of Physics and Astronomy, Northwestern University, Evanston, IL, USA\\
}

\date{Accepted XXX. Received YYY; in original form ZZZ}

\pubyear{2022}

\begin{document}
\label{firstpage}
\pagerange{\pageref{firstpage}--\pageref{lastpage}}
\maketitle

\begin{abstract}

Stripped-envelope supernovae (SESNe) originate from massive stars that lose their envelopes through binary interactions or stellar winds. The connection between SESN subtypes and their progenitors remains poorly understood, as does the influence of initial mass, binarity, explodability, and metallicity on their evolutionary pathways, relative rates, ejecta masses, and progenitor ages. Here, we investigate these properties across a wide metallicity range (0.01–2 $Z_{\odot}$) using \texttt{POSYDON}, a state-of-the-art population synthesis code that incorporates detailed single- and binary-star model grids. We find that the common-envelope channel contributes less than 6\% of SESNe, since unstable mass transfer is found less frequent than previously thought and rarely leads to CE survival when  envelope binding energies are computed from detailed stellar models. The secondary channel accounts for less than 11\%, while the vast majority of SESNe originate from primary stars in binaries undergoing stable mass-transfer episodes. These interactions maintain a largely metallicity-independent SESN parameter space, making the overall SESN rate almost insensitive to metallicity. In contrast, subtype fractions exhibit strong metallicity dependence, though their exact values remain affected by classification thresholds. The age distributions and therefore the progenitor masses of different SESN types also vary significantly with metallicity, revealing metallicity-dependent trends that can be tested observationally. Predicted SESN ejecta masses remain nearly constant across metallicity, in contrast to single-star models, and fall within observed ranges. Future transient surveys, combined with statistical environmental studies that constrain metallicity dependence, will provide decisive tests of these predictions and of the dominant role of binary interactions in shaping SESNe.

\end{abstract}




\section{Introduction}\label{sec:intro}

Core-collapse (CC) supernovae (SNe) are the explosive fate of massive stars. However, there is growing evidence that not all massive stars produce observable energetic transient events \citep[e.g.,][]{2015Smartt, 2015Gerke, 2024De}.  At least for some SNe,
formation of a black hole (BH) makes it difficult to observe the
events \citep[e.g.,][]{2015Gerke, 2015Smartt}. Among those that we observe, their classification relies on the presence/absence of some spectroscopic features \citep[e.g.,][]{1997Filippenko}. Some of these explosions are observed to be hydrogen-deficient, showing weak or no hydrogen lines in their spectra, and are classified as stripped envelope SNe (SESNe). In this class, we find Types IIb, Ib, Ic,  as well as broad-lined Ic (Ic-BL), Ibn and Icn events, which are thought to arise from massive stars that have lost most or all of their hydrogen-rich envelopes due to some mechanism.

Two main  progenitor scenarios have been suggested for SESNe: (i) single, massive Wolf–Rayet (WR) stars that shed their outer layers via stellar winds \citep[e.g.,][]{1986Begelman, 1995Woosley}, and (ii) stars in binary systems that lose their envelopes through mass transfer to a companion \citep[e.g.,][]{1992Podsia, 1998Dedonder, 2010Yoon,  2017Yoon,  2013Eldridge, 2017zap,2019Sharvan, 2025Gilkis}. It is still uncertain whether one or both of these channels can fully account for the observed SESN population.

Given that most massive stars are found in binary or multiple systems \citep[e.g.,][]{1961Blaauw, 1969VandenHeuvel, 2012Sana,2013Sana,2014Kobulnicky,2015Dunstall, 2017Almeida,2017Moe, 2022Banyard, 2025Villase}, many are expected to undergo mass exchange with a companion during their lifetimes. This suggests that a large fraction of SESNe likely originate from stars that have experienced binary interaction in the past. The high observed rate of SESNe  \citep[$\sim$30\% of all  CCSNe, e.g.,][]{1999Cappellaro, 2011Smith, 2011Li, 2017Shivvers, 2017Graur, 2020Perley}, their low ejecta masses \citep[e.g.,][]{2011Drout, 2015Taddia, 2016Lyman, 2018Taddia, 2021Barbarino}, and the lack of confirmed WR star detections in pre-explosion images of Type Ib/c SNe \citep[e.g.,][]{ 2009Smartt, 2013Eldridge, 2016VanDyk}, though some exceptions exist \citep{2012Yoon, 2013Eldridge, 2015Tramper}, seem to favor the binary star scenario. The two detections of Type Ib progenitors in pre-explosion images \citep[one is pending confirmation with follow-up observations,][]{2021Kilpatrick} are probable not a WR star, but an evolved low mass helium star resulting from envelope stripping in binary interaction \citep[][]{2013Eldridge, 2014Fremling, 2015Eldridge, 2015Follateli, 2016Folatelli}. Recently, such non-WR stripped helium stars have been identified in both the Small and Large Magellanic Clouds (SMC and LMC, respectively) by \citet{2023Drout}
and \citet{2023Gotbrg}, revealing a critical missing link in the evolutionary pathway leading to SESNe. Direct detections of Type IIb SN progenitors in pre-explosion images also tend to support the binary star scenario, as nearly all observed progenitors align with this channel \citep[e.g.,][]{2004Maund, 2014Folatelli, 2015Follateli, 2015Maeda, 2017Killpatrick}. 

Although the most direct method for constraining SN progenitors is to observe the progenitor star prior to explosion \citep[e.g.,][]{2013Eldridge, 2009sm}, significant effort has been made to constrain the physical nature of the progenitors of SESNe  by studying their environments \citep[see][for a review]{2015Anserson}. If SESNe primarily arise from single stars, the well-established link between pre-SN mass loss and initial stellar mass suggests that SN types should correlate with progenitor mass. In this framework, the most massive stars would more efficiently shed their H and He envelopes than less massive ones, making SESNe progenitors, on average, more massive than those of Type II SNe. 
This leads to a proposed sequence of CCSN types arranged by increasing progenitor mass: SN II → SN IIb → SN Ib → SN Ic.  While some statistical enviromental studies support this sequence of SN types as a function of increasing progenitor mass \citep[e.g.,][]{2014Galbany, 2017Kangas, 2018Kuncarayakti, 2023A&A...677A..28P}, others report significantly different trends \citep[][]{2012Kelly, 2012Anderson, 2015Anserson, 2024Solar}, suggesting that
our understanding of the accuracy of this picture and how it relates to progenitor mass or other characteristics such as binarity or metallicity is far from complete. 

Metallicity plays also a crucial role in the evolution of massive stars and the resulting SN types. At high metallicities, stars have stronger line-driven winds, allowing for the stripping of the envelope \citep[][]{1975Castor, 2017Renzo, 2024Josiek}.  This would lead to an increasing number of H-poor SNe relative to H-rich ones, and if stellar winds were the main factor for envelope stripping, a clear metallicity trend in the SESN to Type II (H-rich ones) SN ratio would be expected,  with this ratio decreasing at lower metallicities. However, recent studies indicate that metallicity has a minor impact on SESN occurrence. Some observations indicate that the SESN-to-Type II SN ratio remains nearly constant across metallicities, although the coverage is limited to above $\sim $0.4 $\rm Z_{\odot}$ because the samples are drawn from targeted surveys that preferentially select bright galaxies \citep[e.g.,][]{2012Kelly, 2018Kuncarayakti}, a result that contradicts the predictions of single-star evolution models.  Furthermore, recent work by \citet{2024Xi} finds that metallicity has only a minor effect on SESN production within 12 + log(O/H) = 8.1–8.7 dex. These findings suggest that metallicity-insensitive processes, especially binary interactions, are likely the dominant drivers of SESNe, rather than single-star evolution.

Metallicity should also affect the SESN production by binary progenitor models,  in a way that higher metallicity would still produce more highly stripped progenitors \citep[e.g.,][]{2017Eldridge, 2017Yoon}. At lower metallicities, binary interactions are anticipated to result in fewer H-poor envelope SNe. This is because reduced internal opacity, altered nuclear reaction rates, and variations in mean molecular weight resulting from different initial chemical compositions, affect the evolution of stellar radii as well as the timing and evolutionary stage at which the donor initiates mass transfer,  allowing stars to contract within their Roche lobe before the H envelope is fully removed \citep[e.g.,][]{2017Gotberg,2020Laplace, 2020Klencki,2022Klencki, 2022Xin}. As a result, a significant amount of the H-envelope can be retained, which cannot be efficiently removed due to weaker stellar winds \citep[e.g.,][]{2001Vink, 2019Gilkis, 2017Yoon, 2017Gotberg}. However, \cite{2020Laplace}, found that stars that have  experienced roche lobe overflow (RLOF)  at lower metallicities can retain a H-rich envelope and can, in principle, expand to giant
sizes (>400 R$\odot$) for all masses (with initial masses ($M_{\rm ZAMS}$) ranging from 8.9 to 15 $\rm M_{\odot}$, where ZAMS stands for zero-age main sequence) considered in their work, suggesting that they can possibly fill their Roche lobe anew initiating another mass transfer. This would have important consequences on the resulting SN types as well as the ejecta mass during SNe. To account for the effects of metallicity and the effects that a leftover hydrogen may have on the properties and evolution of partially-stripped stars, and subsequently on SESNe, it is essential to perform detailed binary evolution calculations that self-consistently model the outcomes of Roche lobe stripping and taking into account all the possible mass transfer episodes.

The majority of binary population synthesis codes  such as \texttt{BSE} \citep[][]{2002Hurley}, \texttt{Binary\_c} \citep[][]{2004Izzard,2006Izzard}, \texttt{COMPAS} \citep[e.g.,][]{2017Stevenson,2018Vigna}, \texttt{COSMIC} \citep[][]{2020Breivik} and \texttt{SEVN} \citep[][]{2015Mapeli, 2019Mapeli} 
are based on fitting formulae or interpolation in grids of models of single hydrogen-rich or helium stars, without computing self-consistently  the binary stripping process. 
As a result, they have limited power in capturing the impact of a left-over layer of hydrogen on the surface on the properties of stripped stars, an effect that becomes even more important in stripping at lower metallicities. 
In addition, most of these codes apply simplified criteria for determining stellar explodability, typically without considering the detailed internal structure of the progenitor stars, although some exceptions exist \citep[e.g., \texttt{SEVN},][]{2015Mapeli}. 
On the other hand, detailed or semi-detailed stellar evolution simulations have been used either for specific CCSN types or for a relatively small number of progenitor models, typically covering only a narrow range of metallicities \citep[e.g.,][]{2010Yoon, 2015Yoon,2017Eldridge,2018stanway, 2017Yoon, 2019Sharvan, 2024Ercolino, 2025Gilkis}.

Motivated by these open questions and the limitations of current theoretical models, this study presents a multi-metallicity analysis using the binary population synthesis code \texttt{POSYDON} \citep[][]{2023Fragos, 2024Andrews}, which is built upon extensive grids of binary star simulations generated with \texttt{MESA}. By modeling the evolution of both primary and secondary stars, accounting for the most plausible mass transfer scenarios and evolutionary pathways, and incorporating updated SN prescriptions based on the detailed internal structure of progenitor stars, our goal is to investigate how the typical progenitor masses, relative rates, and ejecta masses of different SESN subtypes vary with metallicity. This analysis enables us to explore the role of binary evolution in the formation of SESNe across a range of metallicities and to disentangle the complex interplay among key factors such as metallicity, binarity, and explodability that shape the evolutionary pathways leading to various SESN types.

The structure of this paper is as follows: Section \ref{sec:2} outlines our model assumptions, methodology, and the criteria used for SN classification. In Section \ref{sec:3}, we present a parameter survey along with the evolutionary pathways of SESN progenitors at solar metallicity. Section \ref{sec: metallicity} explores the impact of metallicity on SESN types and the corresponding progenitor evolution channels. In Section \ref{sec:5}, we present our population synthesis results, including the relative likelihood of different evolutionary channels, relative rates, ejecta masses, explosion times of different SESNe types as a function of metallicity, and comparisons with observational data. Section \ref{sec:6} discusses the implications, uncertainties, and variations of our findings. Finally, Section \ref{sec:7} summarizes our conclusions.



\section{Methodology}\label{sec:2}

In this work, we used the Version 2 of the binary population synthesis code \texttt{POSYDON} \footnote{\href{https://posydon.org/}{posydon.org}. Exact commit used in this work is \href{https://github.com/POSYDON-code/POSYDON/tree/Dimitris_WD_mergers}{891c5897}.}\citep{2023Fragos, 2024Andrews}, which, in contrast to Version 1, models the evolution of stellar binaries across a cosmological range of metallicities and incorporates a treatment for stellar mergers. \texttt{POSYDON}\footnote{$\sim$Data Release 2 grids are archived in the POSYDON Zenodo community at  \href{https://zenodo.org/records/15194708}{https://zenodo.org/records/15194708}}, which is open source and publicly available, uses extensive detailed single- and binary-star model grids, calculated with the \texttt{MESA} structure and binary evolution code \citep{2011Paxton,2013Paxton, 2015Paxton, 2018Paxton, 2019Paxton, 2023Jermyn}  to self-consistently follow the entire evolution of both stars in a stellar binary. For a detailed description of the code and assumptions of stellar and binary evolution physics, see \cite{2023Fragos} and \cite{2024Andrews}. The same computational set up has been implemented in a recent study of the population of binary companions next to SESNe  \citep{2025bZapartas}. Here, we present a brief overview of the key parameters and fundamental physical assumptions used in our population synthesis.

\textit{Initializing our binary populations:} We evolved  2 $\times~ 10^{5}$ stellar systems for each of six different metallicities (2, 1, 0.45, 0.2, 0.1, 0.01) $Z_{\odot}$ \citep[with $\rm Z_{\odot}$ = 0.0142;][]{2009Asplund} adopting a binary fraction ($f_{\rm bin}$) of 0.6. This value represents a reasonable average drawn from various spectroscopic studies, which consistently show that more than 50\% of O- and B-type stars in the Galaxy and the LMC are found in binary or multiple systems within the orbital period range relevant for interactions (< 3500 days) \cite[e.g.,][]{2012Sana, 2013Sana,2014Kobulnicky, 2015Dunstall,2017Almeida, 2021Villase,2022Banyard, 2025Villase}.
The distribution of orbital periods is remarkably similar across most of these samples \citep[][]{2017Almeida, 2021Villase, 2022Banyard}, suggesting
that O- and B-type stars in multiple systems are formed with
similar efficiency and orbital configurations within the metallicity ranges studied so far. However, more recent studies point to a possible anti-correlation between metallicity and the close (binaries at $\rm log_{10}(\rm P_{orb}/days)$ < 3.5) binary fraction among B-type stars, with the SMC ( Z $\sim 0.2 ~\rm Z_{\odot}$) exhibiting significantly higher multiplicity fractions, up to 80\%, compared to the LMC and the Galaxy \citep[][]{2025Villase}. This potential enhancement in close binaries at low metallicity may have important consequences for the evolution of massive stars and the production of SESNe. Therefore, while we adopt $f_{\rm bin}$=0.6 as a representative average based on the bulk of existing data, we also explore higher binary fractions of  0.8 to capture the possible increase in close binary incidence at lower metallicities in subsection \ref{6.2}.

We apply a burst star formation history, which assumes all stars are formed at the same time, and evolve our binaries for 13.8 Gyr or until both stars have completed their evolution and transitioned into compact objects. For primaries (initially more massive stars) and single stars, we cover the same initial-mass range 3.94-250 $\rm M_{\odot}$. The initial binary component masses and orbital configurations are drawn from observationally motivated distributions, with primary and single star masses sampled using a Kroupa power-law initial mass function (IMF) \citep{2001Kroupa}. Initial orbital periods are drawn from the distribution of \citet[][]{2013Sana} for systems with initial primary masses greater than 15 $\rm M_{\odot}$ and orbital periods ranging from 1.4 to 3170 days. Since the \citet[][]{2013Sana} distribution is undefined for periods shorter than 1.4 days, we extend it by assuming a distribution that is uniform in $\log_{10} P_{\rm orb,i}$ down to a minimum period of 0.75 days, following \citet{2023Fragos}. For systems with initial primary masses below 15 $\rm M_{\odot}$, we adopt a logarithmically uniform distribution across the entire period range. 
For the binaries, secondary (initially less massive stars) masses are determined by sampling uniformly in the mass ratio $q = M_{2}/M_{1}$ in the bounds [0, 1] \citep[][]{2013Sana} with a minimum allowed mass of 0.35 $\rm M_{\odot}$. All binaries start with circular orbits and stellar spins aligned with the orbital period, indicating effective tidal synchronization during the pre-main-sequence phase. In all population models considered in this study, we used a nearest-neighbor method, assigning each system to the closest \texttt{MESA} model in the mass–q–period space of the relevant grid, without performing interpolation within the grids \citep[see Section 7 in][]{2023Fragos}. To assess the sensitivity of our results to this choice, we also tested an interpolation-based approach, described in subsection~\ref{6.2}.
 
\textit{Stellar winds and their metallicity dependence:} The models adopt the default wind prescriptions, primarily following the \texttt{MESA} "Dutch" scheme. This includes the winds of \citet[][]{1988deJager} for cool H-rich stars and \citet[][]{2000Vink, 2001Vink} for hot H-rich stars. For stars with a surface H mass fraction below 0.4, the winds are replaced by the WR winds of \citet[][]{2000Nugis}. For cool stars (effective temperature below 8000 K) with initial masses below 8 $\rm M_\odot$, we also adopt the \citet[][]{reimers1975} and \citet[][]{1995Blocker} wind prescriptions, appropriately scaled for stars on the first ascent of the giant branch and for those in the thermally pulsing asymptotic giant branch (TP-AGB) phase \citep[see details in][]{2024Andrews}. 

Both the \citet[][]{2000Vink,2001Vink} and \citet[][]{2000Nugis} prescriptions include an explicit dependence on metallicity. However, for winds from cooler red (super)giants and AGB stars \citep[][]{1988deJager, reimers1975, 1995Blocker},the metallicity dependence is uncertain and likely weak. This is supported both empirically \citep[e.g.,][]{2005vanloon, 2012Groenewegen, 2017Goldman, 2024Antonias} and theoretically \citep[][]{2021Kee}. Following the approach used in \texttt{MIST} models \citep{2016Choi}, we therefore assume these winds to be metallicity-independent.

In addition, we include a simple prescription for luminous blue variable (LBV)-type winds, following \citet[][]{2010Belczynski}. Specifically, we impose an enhanced mass-loss rate of $10^{-4} ~\rm M_\odot,\text{yr}^{-1}$ for stars that exceed the Humphreys–Davidson limit \citep{1979HUmphreys}, defined here as stars with $L > 6 \times 10^5 \rm L_\odot$ and $(R/\rm R_\odot)~ \times~$$(L/\rm L_\odot)^{1/2} > 10^5$. Given the significant theoretical and observational uncertainties surrounding LBV winds and their potential metallicity dependence, and the simplicity of our adopted prescription, we do not account for any metallicity scaling in the LBV-like wind model. A discussion on the possible
impact of different wind prescriptions on our results can be found in Section \ref{sec: variations}.

\textit{Mass transfer:} The \texttt{POSYDON} grids employed in this study are based on detailed \texttt{MESA} binary simulations, enabling a self-consistent determination of mass transfer rates via RLOF and their effects on both the donor and accretor structures.  Accretion onto non-degenerate stars is treated as conservative until the accretor approaches break-up rotation velocities \citep[e.g.,][]{1981Packet, 1998Langer}, whereas accretion onto degenerate stars is Eddington-limited. A comprehensive description of \texttt{POSYDON's} mass transfer treatment is provided in Section 4.2 of \citet[][]{2023Fragos}.

\textit{Common envelope:} 
In \texttt{POSYDON}, a binary system is assumed to enter dynamically unstable mass transfer when any of the following three criteria are met: (i) the donor's mass-transfer rate exceeds 0.1 $\rm M_{\odot}$ yr$^{-1}$, (ii) RLOF occurs through the outer $L_{2}$ Lagrangian point, or (iii) the photon trapping radius of the degenerate accretor becomes comparable to its Roche lobe radius (see subsection 4.2 in \citet[][]{2023Fragos} for more details). When any of these conditions are satisfied, the system enters a common envelope (CE) phase, which can result in either envelope ejection or a merger. Our default model calculates the orbital evolution of a binary system through CE evolution using a two-step approach. The first step employs the $\rm \alpha_{CE}$–$\rm \lambda_{CE}$ formalism \citep[][]{1984Webbink, 1988Livio}, which assumes that a fraction $\rm \alpha_{CE}$ of the orbital energy released during the spiral-in is used to unbind the envelope. In our fiducial models, we adopt $\rm \alpha_{CE} = 1$, consistent with earlier population synthesis studies \citep[e.g.,][]{2002Hurley}. However, $\rm \alpha_{CE}$ is generally a free parameter in \texttt{POSYDON}, and we explore alternative values in subsection \ref{6.2}.

The parameter $\rm \lambda_{CE}$, which characterizes the binding energy of the envelope and thus determines whether a binary merges during CE evolution, is computed from the detailed stellar structure of the donor at the onset of the CE phase or of both stars in the case of a double CE. The envelope binding energy, and therefore the CE outcome, is sensitive to the chosen core-envelope boundary, as the deeper layers of the envelope are more tightly bound \citep[][]{2000Dewi, 2013Ivanova, 2019Fragos}. To account for this, we consider different boundary definitions: for H-rich stars, the default boundary is where the hydrogen mass fraction drops below 0.3, though we also examine a threshold of 0.1 (see subsection \ref{6.2}); for stripped stars, the boundary is defined where the combined H and He mass fraction falls below 0.1.

Following arguments from \citet{2011Ivanova} and simulations by \citet{2019Fragos} and \citet{2021Marchant}, we assume the system detaches, if sufficient orbital energy is available, leaving a thin residual envelope surrounding the core. A second phase of stable, non-conservative mass transfer is then assumed to occur, during which the donor is stripped down to its 1\% H-mass fraction boundary, assuming that the thin envelope is lost from the system, carrying the specific angular momentum of the accretor. Provided that neither star overfills its Roche lobe during these phases, the binary survives the CE phase, albeit with a much smaller orbital separation. The exposed core of the donor is then treated as a helium star in subsequent evolution.

\textit{Mergers:} In the case of mergers, we model the evolution of the coalesced product following the method described by \citet[][]{2024Andrews}.  For systems where the donor is a post-main-sequence (post-MS) star with a main-sequence (MS) companion or when both stars are post-MS, the two stellar cores spiral into each other and the merger outcome depends on their relative densities. Layers with similar dominant chemical compositions are assumed to mix completely, with final abundances determined by the mass-weighted average of each star’s corresponding layer \citep[see details in][]{2024Andrews}. We approximate the merger product with a single star with He core mass equal to the sum of the He core masses of both pre-merger stars (for MS stars, the He core mass is considered zero) initialized with a composition that is the mass-weighted average of the compositions in the pre-merger cores, and an envelope that remains primordial in composition (not trying to match the total mass as in default \citet[][]{2024Andrews}). This approach enables a more reliable estimate of whether and when the system will undergo a SN explosion. However we stress that multi-dimensional hydrodynamic simulations show that mixing plays a crucial role in stellar mergers \citep[e.g.,][]{2002Lombardi, 2002Ivanova, 2013Glebbeek, 2019Schneider, 2020Renzo_merger, 2025Patton}. This mixing can expand the effective core size in mergers of two MS stars, boosting the effects of rejuvenation, while in mergers involving post-MS stars, it can lead to smaller cores. Moreover,
nuclear-processed material from both stars may be redistributed into the outer layers of the merger remnant, a process not incorporated in
our models. As a result, both the chemical surface enrichment and core properties observed in such simulations are not fully captured by our models. These differences in core properties could potentially
affect stellar explodability, primarily impacting Type II SNe, though this remains uncertain and warrants further investigation.

\textit{Supernovae and compact object formation:} 
We simulate SN explosions and compact object formation using the \cite{2020Patton} model for iron CCSNe and the \citet[][]{2015Tauris} model for electron-capture SNe (ECSNe). The ECSNe model applies to stars that do not ignite oxygen and  have carbon-oxygen (C/O) core masses in the range  1.37 $\rm M_{\odot} \le$ $M_{\rm C/O-core} \le$ 1.43 $\rm M_{\odot}$. Stars with core masses below the lower limit for ECSN evolve into white dwarfs (WDs).  In the \cite{2020Patton} prescription, the explodability of a pre-SN core is determined based on the C/O core mass and the average core carbon abundance at carbon ignition. For comparison, we also explore two alternative engines, the
N20 calibrated scheme of  \citet[][]{2016Sukhbold} and the  
delayed explosion model of \citet[][]{2012Fryer}, as discussed
in subsection \ref{6.2}. 

For each single- and binary-star model in our grids, we record these two values and use a k-nearest neighbor interpolation (k = 5) to map them to the explodability parameters $M_{4}$ and $\mu_{4}$ from \cite{2016Ertl}, following the approach described in \cite{2020Patton}. These parameters allow us to determine whether an SN is successful. If successful, the resulting neutron star (NS) mass is estimated to be approximately equal to $M_{4}$. BHs are assumed to form only from failed explosions that result in direct collapse. For the \cite{2020Patton} model, we use the option with the SN engine N20, incorporating the updated calibration from \cite{2020Ertl}. NSs receive natal kicks from a Maxwellian distribution, with dispersions of 265 km/s for CCSNe \citep{2005Hobbs} and 20 km/s for ECSNe \citep{2019Giacobbo}. BHs receive natal kicks with the same dispersion $\sigma_{\rm CCSNe}$, scaled by the BH mass ($M_{\rm BH}$) with 1.4 $\rm M_{\odot}$/$M_{\rm BH}$, following \cite{2023Fragos}. Another parameter that we investigate in subsection \ref{6.2} is the impact of SN kicks. Some studies have implied that the NS kick should be weaker. For example, \citet[][]{2023DOHERTY} obtained a velocity dispersion of $\sigma_{\rm CCSN}$ = 61.6 $\rm  km s^{-1}$
from the populations of NS binaries. It should be considered
as the lower limit because their sample excludes the disrupted systems where the NSs received large kicks. As a result, we repeated our calculations considering $\sigma_{\rm CCSN}$ = 61.6 $\rm  km s^{-1}$ for CCSN at solar and 0.1 $\rm Z_{\odot}$ metallicity (see subsection \ref{6.2}). BH kicks follow the same distribution but rescaled by a factor of 1.4 $\rm M_{\rm \odot}$/$ M_{\rm BH}$ as previously. If the SN kick does not disrupt the binary, it may induce orbital eccentricity. For subsequent mass transfer episodes, we assume instantaneous circularization of the orbit at the time of RLOF. Specifically, when RLOF occurs, \texttt{POSYDON} circularizes the orbit at periastron.

\subsection{SN Classification \& model definitions}\label{2.1}

To classify pre-CC models as progenitors of SNe II, IIb, Ib, or Ic, we use the total H mass in the ejecta ($M_{\rm H, ej}$) as a primary indicator, along with the pre-SN surface $^{4}\mathrm{He}$ and $^{14}\mathrm{N}$ abundances evaluated at the time of core carbon depletion. However, the relationship between the structure of the pre-SN progenitor and the spectroscopic characteristics of the explosion remains an active area of research \citep[e.g.,][]{2018Dessart}. 

For Type IIb progenitors, the adopted criterion was defined as   $M_{\rm H,ej}$ $\leq$ 0.5 $\rm M_{\odot}$ to ensure consistency with the H-envelope masses inferred for SNe IIb, both with and without  detected progenitors \citep[e.g.,][]{1994Woosley, 1996Houck, 2012Bersten, 2014Morales-Garoffolo, 2014Bufano, 2015Morales, 2017Arcavi, 2018Bersten, 2019Fremling}.  Consequently, progenitors with  $M_{\rm H,ej}$  > 0.5 $\rm M_{\odot}$ are classified as Type~II progenitors. 

For Type Ib SNe, which lack hydrogen features in their spectra, we assume a $M_{\rm H,ej}$ $\le$ 0.001 $\rm M_{\odot}$, following the criterion proposed by \citet[][]{2011Dessart}. However, the minimum H mass required to produce a Type IIb rather than a Type Ib appearance remains uncertain. Estimates range from 0.001 $\rm M_{\odot}$ to approximately 0.033$\rm M_{\odot}$ \citep[][]{2012Hatchinger}. Pre-SN imaging of Type Ib progenitors tends to support the higher threshold \citep{2022Gilkis2022}, although only two such progenitors have been identified to date, and one case, SN 2019yvr \citep[][]{2021Kilpatrick}, remains unconfirmed pending follow-up observations. Consequently, for H ejecta masses in the range 0.001 $\rm M_{\odot}$<$ M_{\rm H,ej}$<0.033 $\rm M_{\odot}$, we assign a tentative classification named Type I(I)b, allowing us to investigate how adopting the thresholds proposed by \citet[][]{2012Hatchinger} or \citet[][]{2011Dessart} would affect the predicted parameter space and relative rates for the progenitors of Type IIb and Type Ib SNe as a function of metallicity.

Distinguishing further between Type Ib and Type Ic SNe remains particularly challenging, as no definitive boundary exists in terms of progenitor structure \citep[see][for a comprehensive discussion]{2015Yoon}. It is still unknown how much helium can be hidden in Ib/c SNe.  According to \citep[][]{2020Dessart}, the type of SN produced by helium stars is determined not only by their final (or ejecta) mass but also by the mass and chemical composition of their envelope. Specifically, they reported that Type Ic SNe spectra can arise from helium stars if their mass loss is significant enough to strip away most of the helium-nitrogen envelope. According to their models, Type Ic SNe are produced by helium stars with surface $^{4}\mathrm{He}$ abundances ($X_{(\rm  ^{4}\mathrm{He})}$) below 0.5  and  surface nitrogen $X_{(\rm^{14}\mathrm{N})}<10^{-4}$. Therefore, in this study, we use these two conditions as a criterion to distinguish potential Type Ib SN progenitors from potential Type Ic SN progenitors. In our models, we identify systems where they are nitrogen deficient, yet the $X_{(\rm  ^{4}\mathrm{He})}$ ranges from 0.5 to 0.9. 
In these instances, the SN classification remains uncertain (see model he9 in \citet{2020Dessart}), and determining the precise SN type for these cases requires radiative transfer simulations. However, since only a very small fraction of systems fall into this category at all metallicities and assigning them to either Type Ib or Ic would not  affect our overall results, we adopt a conservative approach and classify them as Type Ib SNe. Additionally, we highlight that the spectral characteristics of a Type I SN are influenced by nickel mixing, and it’s possible that nitrogen-enriched stars could lead to Type Ic SNe as well. Therefore, our calculation may underestimate the number of Type Ic SNe. 

We do not attempt to further differentiate H-poor SNe that interact with circumstellar material (CSM) at the time of explosion, such as Type Ibn or Icn events, since our progenitor models are evolved only up to carbon core depletion. Similarly, we do not differentiate between normal Type Ic and broad-lined Type Ic (Ic-BL) SNe, as accurately capturing these events would require incorporating additional physics, such as central engine activity, for which the key mechanisms remain uncertain. A detailed investigation of these specific subtypes is beyond the scope of this work, as their occurrence rates are relatively low compared to the main SESN population \citep[for example the event rate of Type Ibn is estimated to be approximately 1–2 \% of all CCSNe][]{2008Pastorello, 2022Maeda, 2025Ma, 2025Pessi}. Therefore, all H-poor SN subclasses are grouped under the broader SESN categories considered in this study, with further analysis left to future work.

In this section, we categorize also the mass transfer of binary systems into four main cases, based on the evolutionary stage of the donor star at the onset of mass transfer. Case A refers to mass transfer initiated while the donor is still on the MS. Case B corresponds to mass transfer that begins while the donor is crossing the Hertzsprung Gap (HG, the period between core H depletion and the ascent on the giant branch) or during core He burning, prior to core He depletion. Case C involves mass transfer occurring after core helium depletion. Case BB describes mass transfer initiated after core helium depletion, specifically when the surface H mass fraction ($X_{\rm H}$) drops below 0.1. Additionally, we define a mass tranfer case named "contact during MS" for systems in which both stars simultaneously overfill their Roche lobes while still on the MS. For such cases, we model the evolution using \texttt{MESA’s} contact scheme (see \citet[][]{2016Marchant} and \cite{2023Fragos} for details). In systems that undergo multiple  mass transfer phases, we represent the full mass transfer history by listing each phase in sequence, separated by a slash. For example, case A/B indicates a system where mass transfer begins under case A conditions and is later followed by a case B episode involving the same donor star. In cases where, following an initial mass transfer (MT) phase from the donor to the companion, the companion evolves off the main sequence and initiates a new MT phase back onto the original donor (now the companion) before the latter collapses into a compact object, we refer to this as "reverse mass transfer".

\section{Parameter survey and evolutionary channels of stripped envelope Supernovae at Solar Metallicity}\label{sec:3}

\subsection{Stripped envelope supernovae from primary progenitors}\label{sec:3.1}

\begin{figure*}
\includegraphics[trim=0 0 0 0, clip=true,width=\textwidth,angle=0]{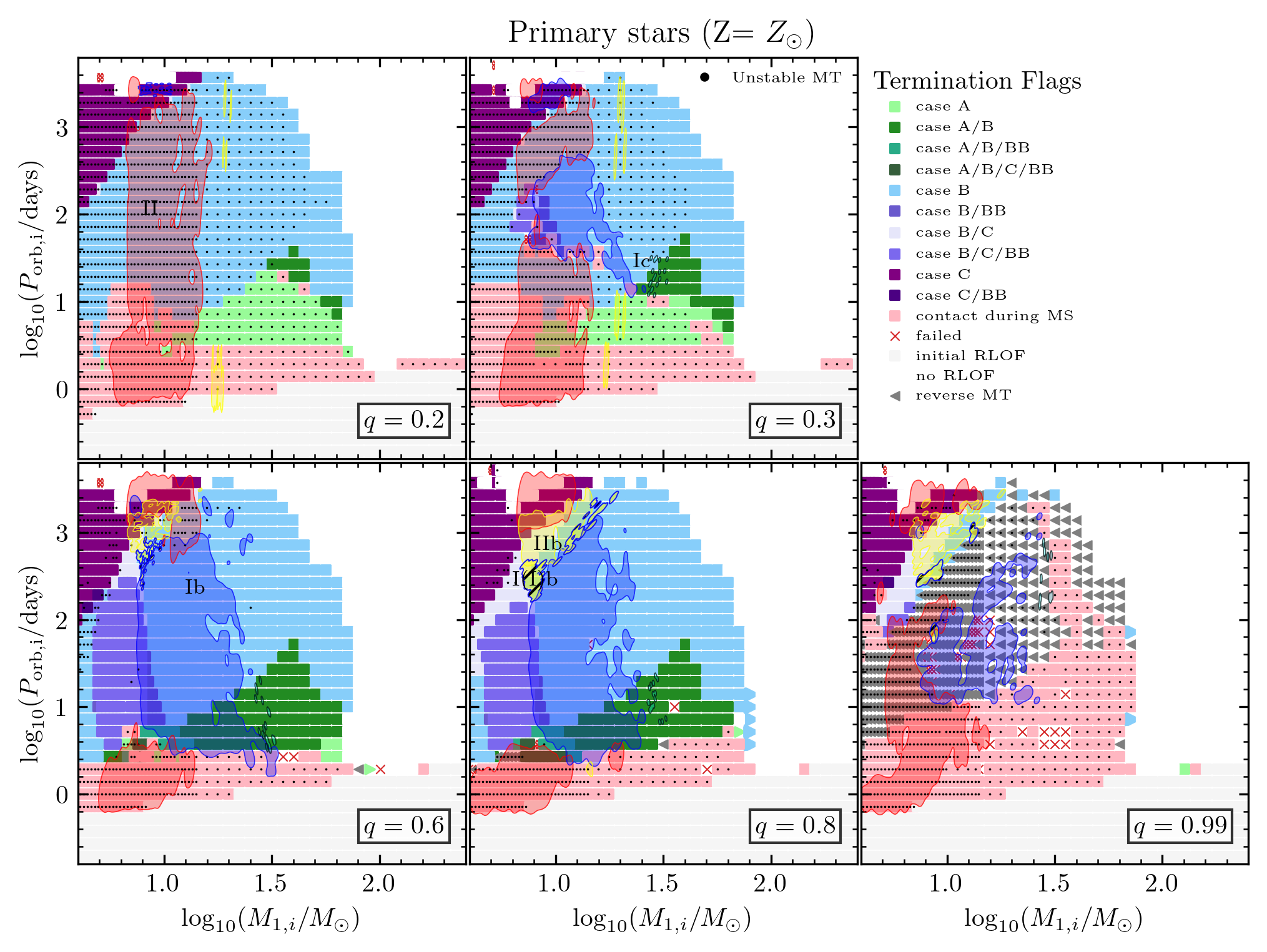} 
\caption {Parameter space for primary progenitors of SN Ic, Ib, I(I)b, IIb and II at solar metallicity. Each panel refers to a different initial mass ratio. See Section \ref{2.1} for SN type classifications and definitions of mass transfer. Here, blue shaded area refer to Type Ib, cyan to Type Ic, yellow to Type IIb, yellow with a blue hatch to Type I(I)b and red to Type II. Each non circular point in the pannels corresponds to one binary simulation with POSYDON, with the marker’s shape and color indicating information about the mass transfer history of the binary till the end of the life of one of the stars.  Pink squares indicate binaries that would overfill their Roche lobes at initialization ("initial RLO"), white area indicates binaries that never overfill their Roche lobes ("no RLO"), colored squares indicate binaries that only the primary star transfer matter to secondary, and colored triangles refers to the binaries that also the secondary transfer material to the primary ("reverse mass transfer"). Black dots refer to binaries that experience unstable MT.} 
\label{fig:primary_solar}
\end{figure*}

To examine how initial mass and orbital period affect binary evolution and SN outcomes, Figure~\ref{fig:primary_solar} shows the predicted final fate of the primary star across a grid of initial masses and orbital periods, at solar metallicity and fixed mass ratios. The SNe are shown as filled, shaded contours, with different colors indicating different SN types (cyan for Type Ic, blue for Type Ib, yellow with a blue hatch for Type I(I)b, yellow for Type IIb, and red for Type II), while non-circular symbols
are used to convey information regarding the binary system's mass transfer history up to the end of the life of one of the stars. This section focuses on the evolutionary pathways and mass transfer processes leading to SESNe, as functions of initial parameters, without assessing the relative contribution of each to specific SN subtypes (see Section \ref{5.1}). 

As shown in all panels, primary stars with initial masses greater than 45 $\rm M_{\odot}$  form BHs and we assume they do not produce a visible SN in electromagnetic waves \citep[although see][]{2018Ott, 2018kuroda, 2022GalYam, 2025Burrows}. Furthermore, primaries that exceed 70 $\rm M_{\odot}$ usually do not interact with their companions.
This is largely due to the fact that very massive stars at high metallicity experience extremely strong stellar winds, which not only widen their orbits but also strip the primary star of its H-rich envelope. As a result, these stars do not expand sufficiently to fill their Roche lobes \citep[see][]{2024Andrews, 2024Kruckow}.  On the other hand, low-mass primary stars with initial masses lower than 8~$\rm M_{\odot}$ at solar metallicity that avoid unstable MT and merging they do not explode as SNe, but die as WDs. This occurs because after case B MT and the subsequent stripping via stellar winds they result in low-mass helium stars with masses lower than  < 3$\rm M_{\odot}$. For such low-helium stars, the expansion of the He
envelope becomes so dramatic during C/O core contraction and/or during core carbon burning, that leads
to another MT phase \citep[e.g.,][]{2017Yoon}. These systems undergoing case B/BB MT, losing so much mass that they can no longer ignite oxygen \citep[e.g.,][]{2010Yoon} and their  C/O core masses stay below 1.37 $\rm M_{\odot}$, which is the minimum threshold for ECSN considered in this study \citep[][]{2015Tauris}.

A comparison between the panels in Figure~\ref{fig:primary_solar} reveals that the initial mass ratio ($q_{i}$) significantly influences MT outcomes, which in turn strongly affect the resulting SN types. For $q_{i}$ $\le$ 0.2, nearly all interacting systems experience unstable MT. Although this typically leads to a merger event and most likely to a type II SN, we identify a narrow region of parameter space, characterized by $M_{1,i}$~$\le$~11~$\rm M_{\odot}$ and $P_{\rm orb,i}$ > 2200 days, where primaries undergoing unstable case~C MT, survive a CE, ultimately producing Type Ib SNe. At such low $q_{i}$, the CE channel represents the only viable pathway for the formation of Type Ib SNe. In contrast, for $0.2 < q_{i} < 0.9$, stable MT becomes more common and the parameter space for binary SNe Ib increases dramatically with increasing $q_{i}$. In this regime,  the largest parameter space for Type Ib SNe at solar metallicity corresponds to case B MT, which occurs across a broad range of initial orbital periods 
and initial primary masses 
depending on the initial mass ratio \citep[][]{1969VandenHeuvel}. This is because in most cases strong winds remove the remaining hydrogen envelope after
detachment.
For relatively short orbital periods (around $P_{\rm orb,i}$ $\lesssim$ 25 days) and primary masses  $M_{1}$$\ge$ 12$ \rm M_{\odot}$, Type Ib SNe originate from primaries that undergo case A/B MT. For lower primary masses and even shorter periods periods ($P_{\rm orb,i}$ $\le$ 10 days) case A/B/C/BB or case A/B/BB MT becomes apparent as only lower primary masses can expand signiﬁcantly after core helium exhaustion. 
For systems with $P_{\rm orb,i}$ $\gtrsim$ 10 days and primary masses below 10~$\rm M_{\odot}$, formation of Type Ib occurs primarily through case B/BB or B/C/BB MT episodes. We also identify a narrow parameter space at very short initial orbital periods ($P_{\rm orb,i}$ $\le$ 3 days) where Type Ib SNe can result from systems undergoing "contact during MS" evolution. However, this channel is more prominent for primaries with $M_{1,i}$>12~$\rm M_{\odot}$, as lower-mass primaries tend to merge during this phase.

For $q_{i}$ greater than 0.9  unstable MT becomes dominant again. This is due to a larger parameter space for contact binaries, where both stars fill their Roche lobes simultaneously while on MS, and cases of reverse MT. As a result, most (though not all) of these systems eventually enter a CE phase. At such high mass ratios, Type Ib SNe are predominantly formed through either stable or unstable reverse MT. In the latter case, the stars enter a (double) CE phase, and if survive the CE can lead to the formation of Type Ib SNe.

For similar regions of the parameter space, particularly those where the initial mass ratio is close to unity, primaries with masses exceeding $25 \rm M_{\odot}$ undergoing unstable reverse MT and surviving the CE can lead to the formation of Type Ic SNe across a wide range of initial orbital periods. 
For lower $q_{i}$, however, Type Ic SNe are confined to systems with the shortest initial orbital periods (around $P_{\rm orb,i}$ $\le$ 30 days) and the most massive primaries ($>20 ~\rm M_{\odot}$) (able to undergo SNe) experiencing case A, case B, or case A/B MT depending on the mass ratio. A notable trend observed in Figure \ref{fig:primary_solar} for Type Ic SNe is that, as the mass ratio decreases (for 0.2 < $q_{i}$ < 0.9),  more massive stars are capable of producing Type Ic SNe. This is attributed to more efficient envelope stripping at lower mass ratios \citep[e.g.,][]{2017Yoon, 2024Ercolino}, which substantially affects the internal structure and leads to reduced final C/O core masses and higher carbon-to-oxygen core abundance ratios \citep[e.g,][]{2021Schneider, 2021Laplace}, conditions that increase the probability of a SN \citep[e.g.,][]{2020Patton}. With the assumed criteria adopted in this study, we find that the initial masses of SNe Ic progenitors are systematically higher than those of SN Ib progenitors.  

Type IIb SNe, at $q_{i}$<0.5, are generally produced by massive primary stars ($M_{1,i}>15~\rm M_{\odot}$)  experiencing unstable MT, where the system does not survive the CE and merge. Conversely, at mass ratios above 0.5, they predominantly arise from binaries in which the primary stars experience stable MT episode(s). The main evolutionary channels for their formation include stable case C, B/C, and case B MT. Case C Type IIb SNe are typically confined to systems with  $M_{1, i} < 10 \rm ~M_{\odot}$ and  orbital periods greater than 600 days. This is due to the fact that only lower primary masses can
expand signiﬁcantly after core helium exhaustion \citep[e.g.,][]{2019Sharvan, 2020Laplace}. Additionally, case C Type IIb SNe are more frequently associated with systems having lower initial mass ratios ($q_{i}$ < 0.8). At higher mass ratios, the dominant formation channels shift toward case B and B/C, with the B/C path being more prevalent at lower primary masses. These channels become dominant at higher $q_{i}$, as the envelope shed through case B MT becomes less efficient for higher initial mass ratio, allowing more primaries to explode as Type IIb rather than Type Ib SNe \citep[e.g.,][]{2017Yoon,2019Sharvan, 2024Ercolino}. 
The influence of less efficient stripping through case B MT increasing $q_{i}$  is also evident in the distribution of Type I(I)b  SNe, where the parameter space of this subclass broadens with increasing $q_{i}$, except at very high $q_{i}$ values $\sim$ 1, where unstable MT processes dominate. 

Our results also show that the largest parameter space for Type II SNe from primary stars corresponds to systems undergoing unstable MT, where the stars typically fail to eject their common envelope and instead merge. Additionally, Type II SNe can arise from stable case~C or non-interacting binaries, though this occurs within a narrow parameter space characterized by high orbital periods (>1000 days) and relatively low-mass primaries ($M_{1,i}$ < 15 $\rm M_{\odot}$), depending on the initial mass ratio. %

\subsection{Stripped envelope supernovae from secondary progenitors}\label{secondaries_se}

\begin{figure*}
\includegraphics[trim=0 0 0 0, clip=true,width=\textwidth,angle=0]{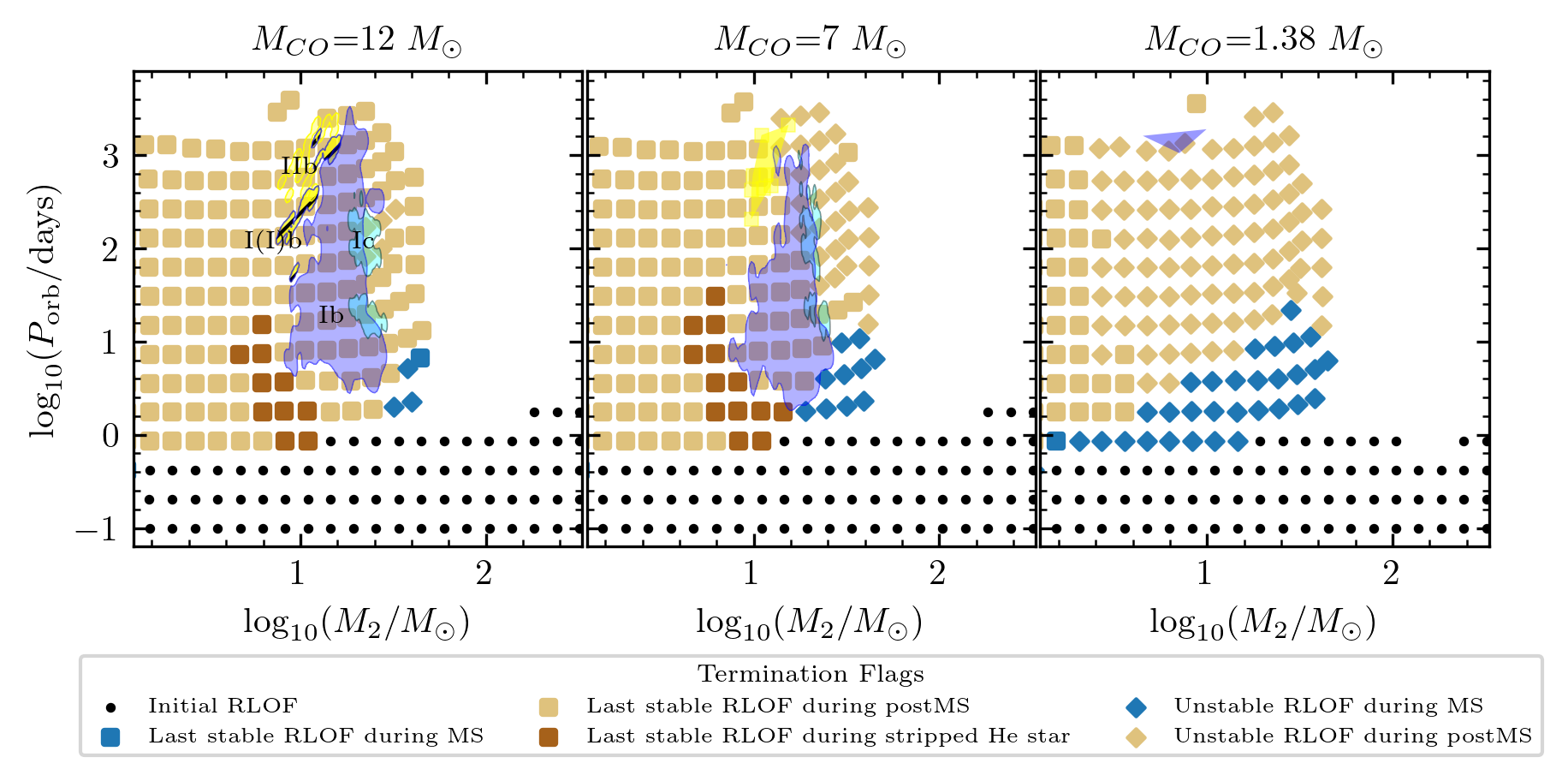} 
\caption {Three 2D slices summarizing the evolution of systems from \texttt{POSYDON} grids of binary star models consisting of a hydrogen-rich
star and a 12 $\rm M_{\odot}$ (left pannel), 7 $\rm M_{\odot}$  (midle pannel) BH and a 1.38 $\rm M_{\odot}$ NS (right pannel) at the onset of Roche lobe overflow. At these panels we oveplot the seconaries that remained bounded following the collapse of the primary star, initiated mass transfer and manage to explode as SNe (contoured shaded areas as Figure \ref{fig:primary_solar}). The different non circular symbols summarize the evolution of each of the models. We distinguish between models that experienced stable (squares) or unstable MT (diamonds). Different colors distinguish the evolutionary phase of the donor/secondary star during the latest episode of MT. Small black dots at low initial periods depict systems that were in initial RLO at birth. Binaries that never initiated MT are not shown here.
} 
\label{fig:sec}
\end{figure*}

Secondaries, like primaries, can contribute to the production of SESNe. However, their overall contribution is expected to be limited, as many are “lost” either through early mergers while both stars are still non-degenerate, occurring in about $\sim$ 56\% of all massive binary systems, or at later stages if the binary survives the primary’s compact object (CO) formation but the secondary eventually merges with the resulting NS (in $\sim$ 67\% of binaries that remain bound to a NS) or BH (in $\sim$1.5\% of binaries that remain bound to a BH).  Another key factor affecting both the SN type and the secondary’s contribution to the total SN population is whether the star becomes disrupted following the collapse of one of the binary components into a CO. We find that disruption is common after the primary’s first collapse: more than $\sim$ 85\% of binaries become unbound due to the natal kick imparted to the newly formed NS, whereas only about 20\% are disrupted by the natal kick associated with BH formation. This is particularly relevant for SESN production at solar metallicity, where we find that disrupted secondaries, if they explode, can produce only Type II or Type IIb SNe, depending on the strength of their stellar winds, thereby restricting their overall contribution to Type Ib and Ic SNe. An important exception arises in systems with high initial mass ratios ($q_{i}$ > 0.9), as demonstrated in subsection \ref{sec:3.1} and illustrated in Figure \ref{fig:a1}, binary interactions, such as stable reverse MT or unstable reverse MT followed by a successful (double) CE ejection, can strip the secondary star before the primary undergoes collapse. This process can lead the secondary to produce a Type Ib or Ic SN regardless of whether the system remains bound or becomes unbound after the collapse of the primary. 

For systems that remain bound after the primary's collapse into a NS, BH, or WD,  \texttt{POSYDON} directs the binaries to the detached binary step where the stars are evolved in detached binaries as essentially single stars, accounting for their effects on the binary’s orbit \citep[see section 8.1 in][]{2023Fragos}. When the secondary star expands during or after the MS  phase, the binary undergoes RLOF. At this point, \texttt{POSYDON} transitions the system into the CO-HMS grid, representing a compact object paired with a hydrogen-rich star. Figure \ref{fig:sec} shows three slices of the grid of \texttt{POSYDON} with different CO masses, $M_{\rm CO}$ = 1.38 $\rm M_{\odot}$ to represent a NS accretor and $M_{\rm CO}$ = 7 $\rm M_{\odot}$  and 12 $\rm M_{\odot}$ to represent an intermediate and a more-massive BH accretor.  The contoured shaded regions in Figure \ref{fig:sec} indicate the predicted final fates of secondaries that remained bound with a compact object and initiate a MT episode. As this grid presents the conditions right before RLOF, any evolution in this grid prior to RLOF is discarded.

Regardless of its evolutionary stage, when a secondary star with a mass $\le$ 25 $\rm M_{\odot}$ overfills its Roche lobe in a binary with a 7–12 $\rm M_{\odot}$ BH companion, the resulting MT is always stable. However, for more massive secondaries, the likelihood of unstable MT increases, particularly when the BH accretor is less massive. As shown in Figure \ref{fig:sec}, Type Ic SNe predominantly originate from secondaries with masses exceeding 25 $\rm M_{\odot}$ that undergo unstable MT and survive the CE phase. The largest region of parameter space corresponds to systems undergoing stable MT onto the BH, which primarily produce Type Ib SNe. In systems with initial orbital periods $\log_{10}(P_{\rm orb}/\rm days) > 2$ and secondary masses below 15 $\rm M_{\odot}$, stable MT results in a Type IIb or Type I(I)b SNe.  For systems that remain bound but the secondary is even more massive (greater than 40 $\rm M_{\odot}$), strong stellar winds prevent it from entering the giant phase and undergoing RLOF. Instead, these secondaries shed their H-rich envelopes, evolve into WR stars, and ultimately form BHs without a SN explosion.

For low mass accretors like NSs, only
the lower mass donors ($M_{\rm 2}$ < 4.5 $\rm M_{\odot}$) do stable MT. For higher mass donors, regardless of the evolutionary state when the
secondary overfills its Roche lobe with a 
1.38 $\rm M_{\odot}$ NS companion, the ensuing MT phase is always unstable due to the large difference in component masses leading to a CE. As shown in Figure \ref{fig:sec}, our models predict that only a limited region of parameter space allows secondaries to survive the CE with a NS and later undergo CCSNe. This outcome is primarily due to the high disruption rate following the first SN explosion: more than $\sim$ 85\% of binaries become unbound as a result of the natal kick imparted to the newly formed NS. Consequently, the surviving  secondary companion typically evolves as an effectively single star. Among the small fraction of systems that remain gravitationally bound, subsequent RLOF from the secondary onto the NS generally resulted in a merger event due to the extreme mass ratio, and thus not producing a SN event.

In cases where the binary system survives the CE, \texttt{POSYDON} transitions the binary to the CO–HeMS grid, which represents binaries composed of a CO and a helium-rich secondary. At this stage, the secondary evolves as a compact, hot helium star. Following core helium exhaustion, the star may expand to form a helium giant. In some systems, depending on the secondary's mass, orbital period, and the mass of the compact object, this expansion triggers a case BB RLOF phase (see Figure 15 in \citet[][]{2024Andrews}). Although we do not display these panels here for brevity, we emphasize that our modeling explicitly incorporates these case BB MT episodes. Thus, the contoured shaded regions indicating the final outcomes of the secondary stars in Figure \ref{fig:sec}, and throughout our analysis, fully reflect their complete evolutionary histories, including these post-helium-burning case BB interactions that may occur after successful CE ejection.

\section{The role of metallicity in stripped envelope supernovae progenitors}\label{sec: metallicity}

\begin{figure*}
\includegraphics[trim=0 0 0 0, clip=true,width=\textwidth,angle=0]{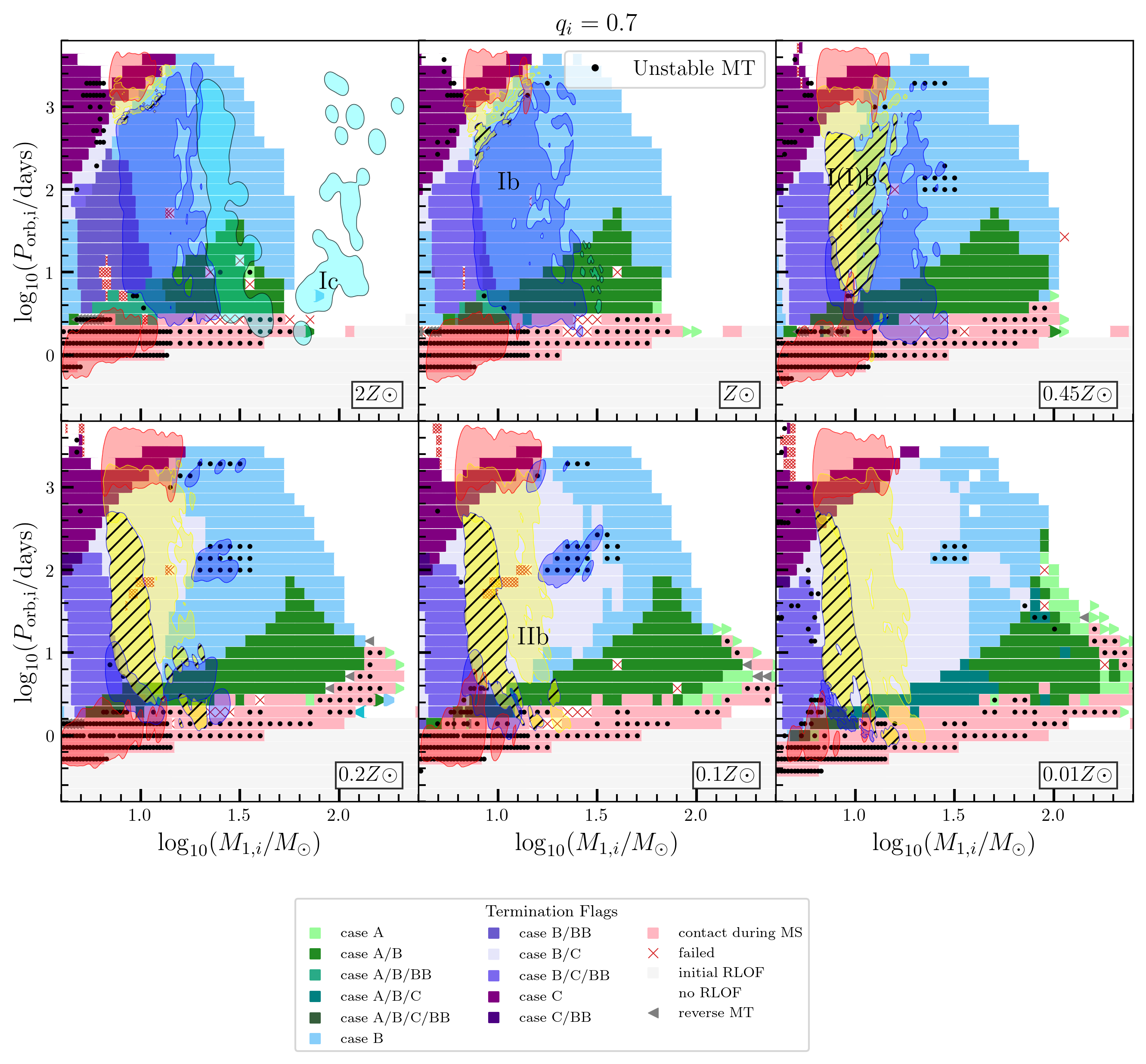} 
\caption {Same as Figure \ref{fig:primary_solar} but for one fixed mass ratio (q=0.7) and as a function of metallicity.} 
\label{fig:MET}
\end{figure*}

To examine the impact of metallicity on the evolutionary pathways of progenitor stars and the resulting SN types, Figure \ref{fig:MET} displays the predicted final fates of primary stars at a representative fixed initial binary mass ratio ($q_{i} = 0.7$), plotted as a function of their initial mass and orbital period for various metallicities. In this section, we primarily focus on primary stars, as secondaries contribute less than 11\% to the total SESNe across all metallicities and predominantly produce Type II SNe (see subsection \ref{5.1} for more details). The metallicity-dependent contribution of secondary stars to each  SN type is provided in Table~\ref{tab:channels}, while the progenitor parameter space at solar and subsolar (0.1 $\rm Z_{\odot}$) metallicity is shown in Figures \ref{fig:a1} and \ref{fig:b2}. 

Comparison between the different panels of Figure \ref{fig:MET} shows MT at this mass ratio is largely stable during post-MS and for lower initial masses ($M_{1,i}$< 30 $\rm M_{\odot}$) the boundaries between no RLO, stable RLO, contact binaries, and initial RLO are all found at similar positions. For lower and higher mass ratios, some variations start to appear (see Appendix \ref{b}), however in general, metallicity plays a minor role in the
MT outcome/stability of low-mass binaries. Of course, as metallicity affects the stellar winds \citep[e.g.,][]{2001Vink, 2019Gilkis}, the star’s opacity  \citep[][]{2017Gotberg}, as well as the evolutionary state of the donor when the first MT takes place  \citep[e.g.,][]{2022Klencki, 2022Xin},  differences in MT cases/history exist. For high-mass stars ($M_{1,i} > 40~{\rm M_{\odot}}$), differences between metallicities become substantial. At low metallicity, even the most massive stars eventually expand, as their weaker stellar winds are often insufficient to fully strip the envelope. Combined with the fact that their orbits widen less, again due to the reduced wind efficiency, they can interact with their companions. However, these stars do not explode as  CCSNe, and we do not explore this regime further in this work.

Metallicity affects not only the explodability of stars, allowing more massive stars to reach the conditions necessary for explosion in higher-metallicity environments, as more efficient envelope stripping through both RLOF and stellar winds leads to the formation of much smaller C/O core masses and higher carbon-to-oxygen core abundance ratios compared to single star counterparts \citep[e.g.,][]{2021Schneider}, but also affects the types of SNe produced and the nature of their progenitors. In Fig. \ref{fig:MET}, a comparison of the panels shows that the parameter space for Type Ic SNe contracts significantly at lower metallicities, with no Type Ic events found below 0.45 $\rm Z_{\odot}$. This indicates that, within the SESNe group (Ib + Ic + IIb), metallicity impacts the production of Type Ic SNe more strongly than it does for Types Ib or IIb, which can form across all metallicities considered in this study. At the highest metallicity explored (2 $\rm Z_{\odot}$), very massive primaries with initial masses between 70 and 200 $\rm M_{\odot}$ in non-interacting binaries are able to produce Type Ic SNe. This provides clear evidence that metallicity-dependent stellar winds play a crucial role in shaping this specific subclass. Type Ic SNe are associated with the most massive progenitors, and the parameter space increases significantly in high-metallicity environments.

On the other hand, the parameter space of SNe IIb and the tentative I(I)b class expands considerably at lower metallicities, though their progenitor properties and evolutionary pathways remain highly metallicity-dependent. In high-metallicity environments ($>0.45~ \rm Z_{\odot}$), SNe IIb and I(I)b primarily originate from low- to intermediate-mass primaries in systems with initial orbital periods exceeding $\sim$100 days (see Section~\ref{sec:3.1}). At lower metallicities ($\leq 0.45~ \rm Z_{\odot}$), SNe IIb tend to arise from more massive progenitors compared to their high-metallicity counterparts. For both SN types, however, these events occupy a much wider range of initial orbital periods and are predominantly produced through case B/C MT (see Figure~\ref{fig:MET}).

At metallicities above 0.45 $\rm Z_{\odot}$, the most massive stars that are able to explode primarily undergo case B MT, after which efficient stellar winds remove the remaining hydrogen, leading to the formation of Type Ib SNe. At lower metallicities (< 0.45 $\rm Z_{\odot}$), where the stripping through both  MT and stellar winds becomes less effective, a significant amount of H is retained at core helium depletion, which can sustain a burning shell, driving the expansion of the envelope after core helium exhaustion \citep[][]{2010Yoon, 2020Laplace}. For the low-metallicity models, H-shell burning dominates the nuclear luminosity around the time of core helium exhaustion and the stars expand shortly after core helium depletion, followed by a contraction and re-expansion phase (see the radial evolution of partially stripped stars after core Helium depletion as a function of metallicity in Appendix \ref{sec:radius evolution}). The first expansion phase is most pronounced in the most massive models ($M_{1,i}$ > 15 $M_{\odot}$), particularly at low metallicities (< 0.2 $\rm Z_{\odot}$) \citep[see also][]{2020Laplace}, as they retain more substantial H envelope masses following the end of case B MT (for example, for a fixed initial mass ratio $q_{i}=0.7$ and an orbital period of 10 days, a 15.7 $\rm M_{\odot}$ donor star retains a H envelope mass of 0.44 $\rm M_{\odot}$ after the cessation of case B mass transfer at 0.01 $\rm Z_{\odot}$ metallicity, whereas an 8.82 $\rm M_{\odot}$ donor retains 0.24 $\rm M_{\odot}$, see Table \ref{tab:properties} for more details). 
More massive stars retain a higher H envelope mass after the first RLOF at low metallicities, mostly because metallicity has a strong effect on what is the typical evolutionary state of donors when they initiate the first MT episode in massive binaries \citep[][]{2020Klencki}. At high (solar-like) metallicities, post-MS interactions almost always occur while the donor is expanding as a HG star. In contrast, at low metallicities, RLOF from core-helium-burning donors occupies a much larger region of parameter space and can even dominate above a certain stellar mass, depending on metallicity \citep[see also][]{2020Klencki}. Unlike HG donors, core-helium-burning stars remain in thermal equilibrium and expand slowly on nuclear timescales. When such a star undergoes MT, its helium core is already close to thermal equilibrium, which facilitates earlier detachment from MT and results in only partial envelope removal. The initial mass threshold at which the donor star begins case B MT during the core helium-burning phase varies with metallicity, but this effect is more pronounced in higher-mass stars. In contrast, lower-mass stars ($M_{1,i}$ $\lesssim$  13 $\rm M_{\odot}$), depending on metallicity and the initial mass ratio initiate MT during HG phase at all metallicities considered in this study, resulting in more efficient envelope stripping through a case B MT.  

Depending on their remaining hydrogen envelope masses after the cessation of the first MT episode, these stars exceed the size of their Roche lobe either during the first expansion phase or during the second expansion phase (or both), where the helium shell burning dominates and initiate a late time MT episode(s). The mass of the retained H layer at core helium depletion influences the maximum radius a star can reach during its first expansion phase \citep[][]{2020Laplace}, thereby increasing the likelihood of Roche lobe filling and the initiation of further mass transfer episodes. These post-helium depletion MT episodes have a significant impact on the SN type, especially at very low metallicities, where the lack of such late-stage interaction would result in notably different SN outcomes. For example, a 15 $\rm M_{\odot}$ donor with an initial mass ratio of  0.7 and an initial orbital period of 100 days at a metallicity of 0.01 $\rm Z_{\odot}$ would retain a hydrogen envelope mass of 0.56 $\rm M_{\odot}$ after case B MT, leading to a Type II SN. However, due to re-expansion of the envelope and a subsequent MT episode, the H envelope mass can be further stripped down to 0.187 $\rm M_{\odot}$, resulting instead in a Type IIb SN (see Appendix \ref{sec:effect_of_initial_orbital_period}, where we demonstrate how this late MT phase affects the SN type across different metallicities and binary configurations). 

Type Ib SNe can be produced from the low mass stars ($M_{1,i}$ < 10 $\rm M_{\odot}$) at all metallicities through case B/BB, B/C/BB or for lower initial orbital periods through cases such as A/B/C/BB or A/B/BB mass transfer.  
The lower remaining H envelope mass after the first MT ceases at low metallicities, compared to the more massive stars \citep[see also][]{2020Klencki} combined with the expansion after core He depletion which is driven either through the H/He shell burning or/and due to the expansion of low helium stars they can produce Type Ib SNe at all metallicities. At high metallicities ($\ge$ 0.45 $Z_{\odot}$), Type Ib SNe can originate from progenitors across a wide mass range, including both high- and low-mass stars. In contrast, at lower metallicities, highly stripped SNe primarily arise from progenitors at the lower end of the mass spectrum considered in this work or from higher mass stars undergoing unstable MT and surviving the CE evolution.  This is illustrated in Figure \ref{fig:MET}, particularly in the panels for 0.45, 0.2, and 0.1 $\rm Z_{\odot}$, where unstable mass transfer leads to Type Ib SNe from progenitors with initial masses in the range of 17–30~$\rm M_{\odot}$ and for a wide range of initial orbital periods.

Having discussed how the parameter space of each subclass of SESNe is influenced by metallicity, we find that the overall parameter space for SESNe remains nearly constant across metallicities. The only change is that stars which explode as one SESN subtype at a given mass and orbital period in one metallicity environment may instead explode as a different SESN subtype in another metallicity environment. A key factor in this stability is the post-core helium-depletion MT phase, which plays a critical role in determining the resulting SN type for both low- and high-mass progenitors. The amount of mass lost during this phase depends on the mass retained at core helium depletion, which influences whether mass transfer is initiated during the first or second expansion phase, the former typically resulting in greater mass loss (see Appendices \ref{sec:radius evolution} and \ref{sec:effect_of_initial_orbital_period}). Consequently, regardless of metallicity or the strength of stellar winds, stars can still be (partially) stripped and explode as SESNe; however, the stellar winds and the efficiency of stripping through RLOF strongly affect the subclasses of SESNe. 

The parameter space of Type II SNe shows little dependence on metallicity. However, the dominant evolutionary channels contributing to this class vary with metallicity.   At low metallicities, stars, including red supergiants, are more compact and exhibit less expansion due to reduced opacities in their outer layers. As a result, very wide binaries are less likely to initiate mass transfer at low metallicity environments, whereas they would at higher metallicities. This leads to an expansion of the parameter space for non-interacting binaries that produce Type II SNe as metallicity decreases  (see Figure \ref{fig:MET}, where the parameter space of "no RLOF" Type II SNe increases with decreasing metallicity). Conversely, the parameter space for mergers occurring during the contact phase decreases with metallicity, since the more compact stellar structure delays the onset of mass transfer to later evolutionary stages, thereby avoiding mergers during the MS contact phase.

\section{Population synthesis results and comparison with observations}\label{sec:5}

\begin{table*}

\caption{Relative frequency of stellar evolution channels contributing to different SN Types as a function of metallicity. These frequencies are normalized to 100\% per SN type and metallicity. For a specific channel, if the system underwent a successful CE ejection, the fraction of such systems is indicated in parentheses (except in the Type II SNe section). If no parentheses are shown, it implies that the CCSNe in that channel did not arise through successful CE evolution. 
In the case of Type II SNe, the parentheses shown in the “Secondary + WD” channel represent the fraction of systems that merge with a white dwarf and the merged product produced a Type II SN \citep[see][for this specific progenitor channel]{2017aZapartas}. }
\label{tab:channels}
\begin{tabular}{|c|c|c|c|c|c|c|}

\multicolumn{7}{|c|}{}                  \\ \hline
CCSN types    &  2 $Z_{\odot}$    & $Z_{\odot}$   & 0.45  $Z_{\odot}$ &  0.2 $Z_{\odot}$ & 0.1 $Z_{\odot}$ & 0.01 $Z_{\odot}$  \\ 
 
 \\ \hline
 Type Ic  & &    &  &  &  &  \\ \hline
 Single & 25.55 \% & - & - & -& -& -\\
 Merged & 2.06 \% &-&  - & -& -& -\\
 Primary & 51.05 \% (0.59 \%) & 39.52 \%  (4.76 \%)& - & -& -& -  \\
 Secondary + BH  & 12.33 \%  (0.46 \%) & 53.57 \% (31.9 \%) & - & -& -& -\\
 Secondary + NS & 0.22 \% & 1.19 \% & - & -& -& -  \\
 Secondary + WD& -& - & - & -& -& -\\
 Secondary (disrupted) & 8.79 \% & 5.72 \% & - & -& -& -\\ \hline
 
 Type Ib  &  &  &    &  &  &    \\  \hline
 Single & - & - & - & - & - & -\\
 Merged & - & 0.06 & 0.03  \%& 0.54  \%& 0.51 \% & 0.75 \% \\
 Primary & 91.97 \% (3.29 \%) & 86.9 (3.06 \%) & 81.7 \% (9.89 \%) & 81.21 \% (33.73 \%) & 85.82 \% (50.1 \%) & 88.01 \% (50 \%)\\
 Secondary + BH & 6.53 \% (0.92 \%) & 8.7 \% (0.2 \%) & 15.85 \% (2 \%)  & 11.7 \% (6.71 \%) & 3.62 \% (3.62 \%) & 4.85 \% (3.42 \%)\\
 Secondary + NS &  0.17 \% (0.04 \%)& 0.11 (0.02 \%) & 0.27 \% (0.1 \%) & 0.2 \%  & 1.28 \% (0.30 \%) & 1.8 \% (0.3 \%)\\
 Secondary + WD & 0.13 \% (0.12 \%) & 0.07 \% (0.07 \%) & 0.05 \% (0.05\%) & - & 0.12 \% (0.12 \%) & 1.91 \% (1.91 \%)\\
 Secondary (disrupted) & 1.2 \% & 1.08 & 2.1 \% & 6.06 \% & 8.29 \% & 2.68 \%\\ \hline

 Type I(I)b  & & &    &  &  &     \\  \hline
Single & -&- &- &- & -& -\\
 Merged & -&- &- & -& -& -\\
 Primary &98.4 \% &97.02 \% &96.48 \% &95.3 \% & 97.14 \% & 98.33 \% \\
 Secondary + BH & 1.6 \% &2.98 \% & 3.52 \% &4.7 \% & 2.86 \% & 1.67 \%  \\
 Secondary + NS &- &- & -& -&- & -\\
 Secondary + WD &- &- & -& -& -& -\\
 Secondary (disrupted) & -&- & -&- &- & -\\ \hline
 Type IIb  & &   &   &  &  &    \\  \hline
 Single & 14.57 \% & 6.57 \% & 9.03 \% & 1.04 \% &- & -\\
 Merged & 9.04 \% & 5.02 \% & 8.35 \% & 0.61 \% & -& -\\
 Primary &  64.63 \%& 85.01 \% & 78.18 \% & 89.57 \% & 90.32 \% & 91.28 \% \\
 Secondary + BH & 1.58 \% & 2.39 \% & 3.07 \%& 8.76 \% & 9.68 \%& 8.72 \% \\
 Secondary + NS & 0.13 \% & -& -& -& -& -\\
 Secondary + WD & 5.75 \% & -& -& -& -& -\\
 Secondary (disrupted)& 5.3 \% & 1.01 \% &1.37 \% & 0.02 \%& -& - \\ \hline

 Type II  &  &  &   &   &  &   \\ %
 \hline
  Single & 45.85 \% & 45.55 \% & 46.21 \% & 46.85 \% & 46.08 \% & 45.77 \%\\
 Merged & 32.37\% & 32.43 \% & 30.48 \%& 29.08 \% & 29.05 \% & 25.08 \%\\
 Primary & 2.12 \%& 2.74 \% & 3.47 \% & 4.23 \% & 5.52 \% & 7.32 \%\\
 Secondary + BH & 0.07 \%&0.12 \% & 0.13 \% & 0.09 \% & 0.1 \% & 0.1 \%\\
 Secondary + NS &  0.02 \%& 0.01 \% & 0.01 \% & 0.02 \%& 0.02 \% & 0.03 \%\\
 Secondary + WD & 1.12 \% (1.09 \%)&1.2 \% (1.1 \%) & 1.32 \% (1.27 \%) & 1.14 \% (1.12 \%)& 1.17 \% (1.1 \%)& 1.09 (1 \%) \\
 Secondary (disrupted) & 18.45 \%& 17.95 \% & 18.38 \% &18.59 \% & 18.06 \% & 20.61 \%\\

 \hline
\end{tabular}
\end{table*}

\begin{figure*}
\centering
\includegraphics[trim=0 0 0 0, clip=true, width=\textwidth, angle=0]{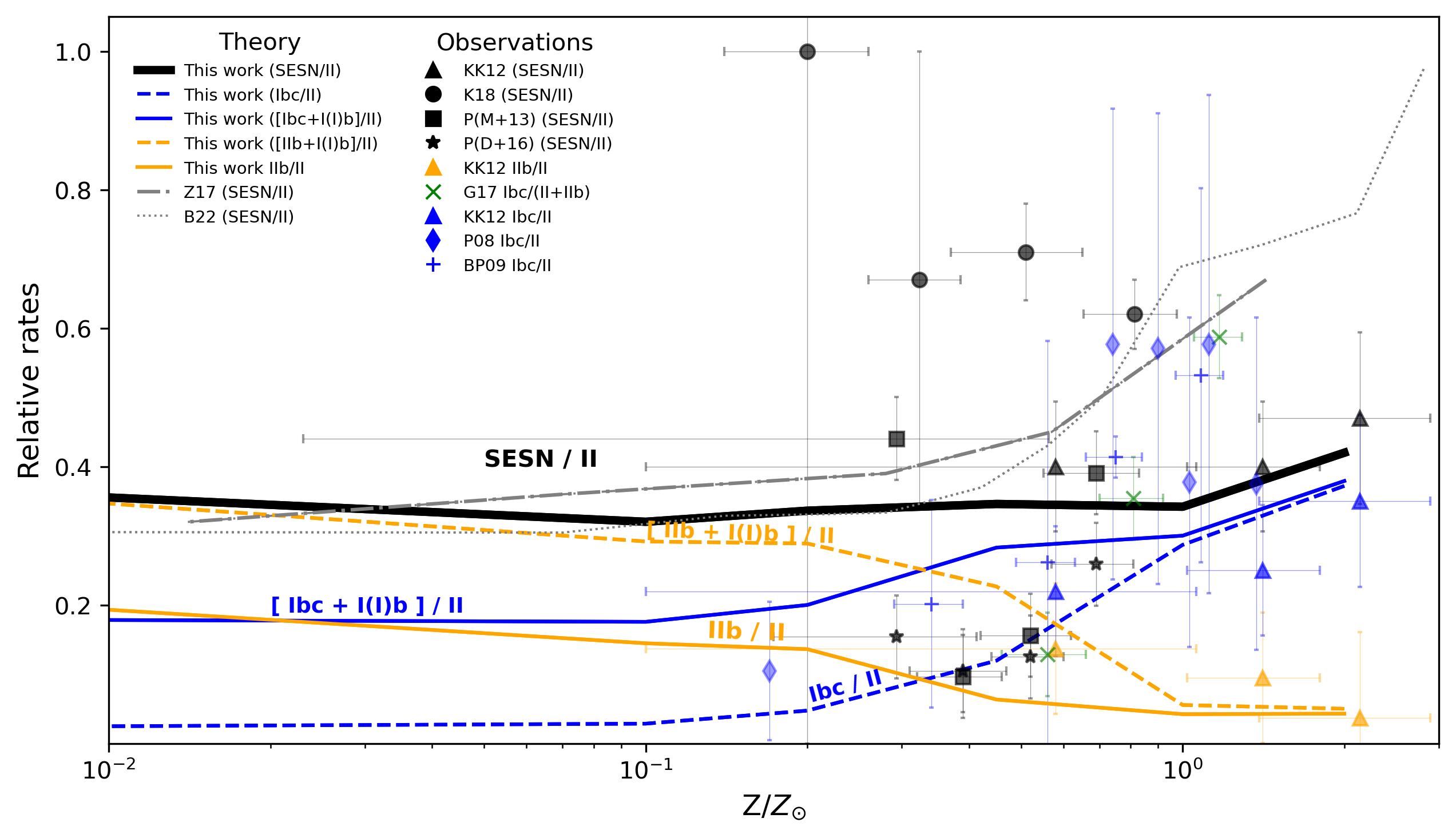} 
\caption{Ratio between different SN types, as a function of metallicity (solid and dashed lines refer to this work, while the different colors refer to ratios between different SN types to Type II) orange: IIb/II, blue: Ibc/II and black (Ic+Ib+IIb)/II. SN observations from the literature, which measured the ratio between the numbers of  different H poor to H rich SNe, are shown as symbols with different colors and shapes regarding the ratio and the study, respectively: P08 \citep[][diamonds]{2008Prieto}; KK12 \citep[][triangles]{2012Kelly}; G17 \citep[]["x"-shaped marker]{2017Graur}; K18 \citep[][circles]{2018Kuncarayakti}, BP09 \citep[][crosses]{2009Boisser}; and P23 \citep[][where M+13, squares, refers to the N2 oxygen abundance index, and D+16, star symbol, refers to the D16 indicator]{2023A&A...677A..28P}. The dotted-dashed black line (SESN/II) represent binary$\_$c models \citep[Z17,][]{2017aZapartas} for a physically motivated population where the majority of stars are in binaries (binary fraction: 70 \%) and assuming that all stars undergo CCSNe. Finally, \texttt{BPASS} (v2: \citep[][]{2016Eldridge, 2016Stanway}) models are represented by a dotted black line \citep[B22,][]{2022Briel}.}
\label{fig:relative_rates}
\end{figure*}

\subsection{Metallicity Dependence of Evolutionary Channel Contributions to Stripped-Envelope Supernovae}\label{5.1}

In Section~\ref{sec:3}, we identified the evolutionary pathways from binary systems that lead to various subclasses of SESNe within our solar-metallicity model, without quantifying the relative contribution of each channel to a specific SN subtype. To address this, and to include both single-star evolution and the effects of metallicity, we classify the progenitor pathways contributing to each SN type into the following seven categories:

\begin{itemize}
    \item \textbf{Single} – SNe resulting from the evolution of isolated stars.
    \item \textbf{Primary} – SNe originating from the evolution of the primary star in a binary system. 
    \item \textbf{Merged} – SNe produced by the evolution of a merger product formed when both stars were non-degenerate at the time of merging.
    \item \textbf{Secondary (disrupted)} – SNe from the evolution of the secondary star after the binary is disrupted by a natal kick.
    \item \textbf{Secondary + WD} – Evolution of the secondary star with a WD companion formed from the primary.
    \item \textbf{Secondary + NS} – SNe that originate from the evolution of the secondary with a NS companion, formed from the primary either via an ECSN  or an iron core-collapse SN.
    \item \textbf{Secondary + BH} – SNe arising from the evolution of the secondary star in a binary system with a BH companion, where the BH was formed from the primary either via direct collapse or a pulsational pair-instability supernova (PPISN), and the system remained bound despite the natal kick.
\end{itemize}

Table \ref{tab:channels} presents the relative frequencies of the various progenitor channels to each SN type across different metallicities. These frequencies are normalized
to 100\% per SN type and metallicity. We present all channels in Table \ref{tab:channels}, but in the text we focus only on the dominant ones, i.e., those contributing $\ge$5\%.

At super-solar metallicity, half of Type Ic SNe originate from the primary channel, while the single-star and secondary channels each contribute approximately 25\%. In contrast, at solar metallicity, Type Ic SNe predominantly arises from secondary stars, which accounts for about 60\% of the cases when all secondary channels are combined. The most common pathway involves a secondary star bound to a BH (53.57\%). In this scenario, Type Ic SN results either from binaries experiencing unstable MT and survive CE (31.9\%), or from stable MT onto the BH (21.67\%). The second most common channel for Type Ic SN formation involves envelope stripping of the primary star through stable MT and stellar winds. The third channel involves secondaries that are ejected following binary disruption caused by a natal kick (8.79\% at super-solar and 5.72\% at solar metallicity). This suggests that these stars are either stripped solely by their stellar winds (predominantly at super-solar metallicity) or have undergone reverse MT prior to the binary’s disruption (i.e., systems with $q_{i} > 0.9$; see Appendix \ref{a}).

For Type Ib SNe, the primary channel contributes more than 80\% across all metallicities. However, as metallicity decreases, both stellar winds and stable MT episodes become less effective at stripping the hydrogen-rich envelope to the extent required for Type Ib SN formation, i.e., reducing the hydrogen mass in the ejecta to below 0.001 $\rm M_{\odot}$. Consequently, CE ejection becomes one of the dominant mechanism for envelope removal in low-metallicity environments (Z $\le$ 0.1 $Z_{\odot}$). This shift is evident in the growing contribution of systems that experience CE ejection to the overall Type Ib population, from only 3.33\% at supersolar metallicity to nearly 50\% at the lowest metallicities considered in this study. This sharp rise primarily results from the small number of systems able to produce Type Ib SNe at low metallicity (see Table \ref{tab:rel_ratio} for SESNe rates across metallicities). We note, however, that although the CE channel becomes important for the production of Type Ib SNe in low-metallicity environments, its total contribution to all SESNe (Types Ic, Ib, I(I)b, and IIb) remains below 6\% across metallicities. This low contribution arises because our detailed binary simulations show that, for primaries in interacting binaries in the 5–40 $~\rm M_{\odot}$ range, where the majority of SN progenitors are expected, stable MT is more common than predicted by rapid \citep[e.g.,][]{2017aZapartas}, or semi-detailed population synthesis models \citep[e.g,][]{2017Eldridge}. Consequently, fewer primaries undergo CE evolution, reducing the likelihood of SESN formation through this channel. Within this mass range, we find that only about 50\% of interacting primaries experience CE evolution, and in most of those cases the systems do not survive the CE phase (computing envelope binding energies from detailed massive stellar models by MESA prior to CE, see Section \ref{sec:2}), a trend that appears largely insensitive to metallicity.

For secondaries, there is no clear monotonic trend with metallicity. The contribution of secondary channels to Type Ib SNe becomes more significant at LMC and SMC metallicities, reaching around 20\%. The dominant channel involves secondaries that remain bound to a BH, undergoing either stable MT or unstable MT followed by successful CE ejection. Additionally, the contribution from the disrupted secondary channel increases from 2 Z$_\odot$ down to 0.1 Z$_\odot$, peaking at 0.1 Z$_\odot$, before decreasing again at lower metallicities. These variations are primarily driven by the sharp decline in the number of primaries producing Type Ib SNe, as many of these systems transition to Type I(I)b or IIb SNe at lower metallicities. This shift significantly alters the relative contributions of the secondary and other channels. Type I(I)b SNe originate almost entirely from the primary channel, contributing over 95\% of cases across all metallicities. The secondary+BH channel contributes only marginally, involving envelope stripping through stable MT from the secondary to a BH.

Across all metallicities, the primary channel remains the dominant contributor to Type IIb SNe, accounting for 64.63\% to 91.28\% of cases, with its relative contribution increasing toward lower metallicities. This trend arises because the single, merged, and disrupted secondary channels, though subdominant, contribute more significantly at higher metallicities (up to approximately 14.6\%, 9\%, and 5.3\%, respectively, at supersolar metallicity), but their influence diminishes or disappears entirely at metallicities below 0.2 Z$_\odot$. This behavior reflects the reduced efficiency of envelope stripping by stellar winds in metal-poor environments. The secondary + BH channel emerges as the most significant alternative pathway, particularly at low metallicities, where its contribution reaches up to 9.68\%.

For Type II SNe, progenitors arise from a mixture of all evolutionary channels across all metallicities. The "single" channel remains the most dominant contributor at each metallicity, followed by the "merged" and "secondary (disrupted)". Together, these three channels account for over 90\% of all Type II SNe across all metallicities.

\subsection{Core collapse Supernova rates as a function of metallicity}\label{5.2}

\begin{table}

\caption{Theoretical relative ratios of various SESNe subtypes to Type II as a function of metallicity}

\label{tab:Modelj}
\begin{tabular}{|c|c|c|c|c|c|}

\multicolumn{6}{|c|}{}                  \\ \hline
Metallicity   &  Ic / II    & Ib / II   & I(I)b / II &  IIb / II   & SESN/II 
 \\ \hline

 2 $\rm Z_{\odot}$  & 7.69 \% &  29.52 \%  &  0.71 \%  &  4.31 \%  & 42.23 \% \\
$\rm Z_{\odot}$ & 0.72 \% &  27.97 \%  &  1.33 \%  &  4.23 \% &  34.25 \%\\
 0.45 $\rm Z_{\odot}$ & -  & 11.97 \% &  16.33 \% & 6.35 \% & 34.65 \%  \\
 0.2 $\rm Z_{\odot}$ & -  &  4.75 \%  & 15.26 \%&  13.62 \% & 33.63 \% \\
  0.1 $\rm Z_{\odot}$ & -   & 2.87 \% & 14.71 \%  & 14.46 \% & 32.03 \% \\
   0.01 $\rm Z_{\odot}$ & -   & 2.49 \%  & 15.34 \% &  19.71 \%  & 37.54 \%\\

\hline
\label{tab:rel_ratio}
\end{tabular}
\end{table}

Table \ref{tab:rel_ratio} and Figure \ref{fig:relative_rates} show the theoretical relative rates of various SN types with respect to Type II SNe for our fiducial population as a function of metallicity. Table \ref{tab:rel_ratio} lists the numerical values, while Figure \ref{fig:relative_rates} illustrates them graphically. The relative rate of SESNe (Ib+Ic+I(I)b+IIb) to Type II SNe remains nearly constant across the full metallicity range. This apparent flatness arises because the metallicity trends of the individual subtypes offset one another: the relative rates of Type Ib/II and Ic/II SNe decrease toward lower metallicities, while those of Type IIb and the tentative class Type I(I)b SNe increase. When combined, these opposing trends yield a net SESN-to-Type II ratio that is effectively independent of metallicity. It is important to emphasize, however, that the relative rates of Type Ib/II and IIb/II depend strongly on the adopted hydrogen-envelope mass threshold in the ejecta, which determines whether an event is classified as Type Ib or Type IIb. The treatment of the tentative class Type I(I)b is particularly crucial, as it substantially affects the inferred subtype rates and can alter the overall results. For example, at $Z = 0.01 ~\rm  Z_{\odot}$, if the tentative Type I(I)b class is assigned to the Type Ib SNe, the relative rate of Type Ib to Type II SNe increases from 2.49\% to 17.83\%. Conversely, if it is assigned to the Type IIb SNe, the relative rate of Type IIb to Type II SNe rises from 19.71\% to 35.05\%. This highligts that the adopted hydrogen-envelope mass threshold distinguishing Type IIb from Type Ib is a critical factor in determining their relative rates, especially in low-metallicity environments. 

Our measured SESNe-to-Type II SN ratios of approximately 34\% and 42\% at $\rm Z_{\odot}$ and $2 \rm Z_{\odot}$, respectively, are broadly consistent with the volume-limited samples of \citet{2011Li} and \citet{2016Graur}\footnote{We quote rates from \citet{2016Graur} using their cut of $3 \times 10^{9}~M_{\odot}$ on host galaxy stellar mass, which separates galaxies below and above the mass of the LMC.}, which report ratios of $\sim$44–50\%. For Type IIb relative to Type II SNe, observations at high metallicities yield $\sim$12–16\% \citep[e.g.,][]{2011Li,2011Smith,2016Graur}, whereas our solar- and supersolar-metallicity models predict values less than half of these rates. Including the tentative Type I(I)b class with Type IIb SNe does not resolve the discrepancy, as the contribution of Type I(I)b in such environments is negligible (see Table~\ref{tab:rel_ratio}). This offset likely arises because our evolutionary models for partially stripped stars adopt the empirical mass-loss prescriptions of \citet{2000Nugis}, which may overestimate the mass-loss rates of low-mass, partially stripped stars (see Section~\ref{6.2} for the role of winds in shaping SN types). At lower metallicities, however, our models (when combining IIb+I(I)b) show better agreement with the implied IIb-to-II ratios from \citet{2016Graur}, who report $\sim$31\%. Since these fractions are derived from volume-limited samples using galaxies spanning a range of luminosity, with most of the CCSN hosts corresponding roughly to metallicities of 0.5–2 $\rm Z_{\odot}$, they ignore metallicity dependence. For this reason, we do not include them in Figure~\ref{fig:relative_rates}. Nonetheless, observational studies consistently demonstrate that the relative rates of different CCSN types depend on host galaxy properties, particularly stellar mass and metallicity, and we include such environment-dependent studies in Figure~\ref{fig:relative_rates} for comparison. 

Numerous SN observations have been used to investigate how the relative rates of  different CCSN types vary
with metallicity. In Figure \ref{fig:relative_rates}, we also include  both  observational measurements from the literature, which report the numbers of various SESNe subclasses relative to Type II SNe, and model predictions of the SESNe-to-Type II ratio as a function of metallicity. We have also included relative rates using metallicity estimates from \citet[][]{2023A&A...677A..28P}, obtained from Integral Field Unit (IFU) observations with the Multi-unit Spectroscopic Explorer \citep[MUSE,][]{2014Muse} instrument. Their sample is composed of CCSNe retrieved from the  All-Sky Automated Survey for SN \citep[ASAS-SN,][]{2014Shappee, 2017Kochanek}, an all-sky untargeted transient survey, thus generating an homogeneous sample as a function of host galaxy properties. The metallicity was extracted from the nearest HII region to each SN, and was estimated using the oxygen abundance using different strong-line methods. Here, we report their metallicity values estimated using the N2 index \citep{2013A&A...559A.114M}, which uses the ratio between [\ion{N}{II}]$\lambda 6584$ and H$\alpha$, and the D16 indicator \citep{2016Ap&SS.361...61D} , which uses the ratio between [\ion{N}{II}]$\lambda 6584$ to [\ion{S}{II}]$\lambda  \lambda  6717,31$ and [\ion{N}{II}]$\lambda 6584$ to H$\alpha$.
The uncertainties in metallicity were estimated from the propagation of line fitting errors from the spectra, while the ratio uncertainties were estimated using Poisson statistics. Despite the high errors owing to the small sample size, the uncertainties in the completeness and metallicity estimations, the observations indicate some general trends with metallicity.

Both observations and our models, regardless of whether the tentative I(I)b class is grouped with Ib or IIb SNe, indicate a general trend: the rate of Type Ib/c SNe relative to Type II SNe decreases with metallicity, while the rate of Type IIb SNe relative to Type II SNe increases toward lower metallicities. Overall, our models are consistent with the observed Ibc-to-II ratios once the observational uncertainties are taken into account. The biggest  difference between our measurements and  observations are those of  \citet{2008Prieto} and \citet[][]{2009Boisser}, who report higher Ibc-to-II ratios at high metallicities; however, their uncertainties are large, and they do not explicitly mention Type IIb SNe, which we, following \citet{2017Graur}, assume were included in the Type II category. Type IIb in the denominator are also included in the work of \citet[][]{2017Graur} so a direct comparison with our theoretical predictions cannot be made. However, they claim in their paper that moving the SNe IIb from the SN II to SESN
column did not have an appreciable effect on their measurements of rate ratio. Our results are consistent with \citet[][]{2017Graur} except the very high metallicities, where they report a higher Ibc/II ratio. Our model predictions are in good agreement with the measurements of \citet{2012Kelly} for both the Ibc-to-II and IIb-to-II ratios. The \citet{2012Kelly} sample is regarded as complete, with metallicities estimated by selecting Sloan Digital Sky Survey (SDSS) fibers closest to the SN explosion sites.
We do not include the surveys of \citet[][]{2012Arcavi}, where the Ibc-to-II ratios were measured with Type IIb SNe classified as Type II. \citet[][]{2012Arcavi} reported these ratios as a function of host-galaxy luminosity, but no significant trend was found. Although we do not plot these measurements in Figure \ref{fig:relative_rates}, we note that their reported range of $\sim$0.2–0.4 is broadly consistent with the other measurements shown. We also exclude the survey of \citet[][]{2015Anserson}, as their sample was assembled from multiple studies that attempted to measure metallicities at SN explosion sites and was therefore biased toward stripped-envelope SNe. As a result, their measurements are inconsistent with the other datasets in Figure \ref{fig:relative_rates}, with number ratios  are biased to
more SESNe rather than SNe II.

For  SESNe (Types Ic+Ib+I(I)b+IIb) relative to Type II SNe, both the observations of \citet{2012Kelly} and \citet{2018Kuncarayakti}, as well as our models, suggest a flat dependence on metallicity. This trend has recently been pointed out by \citet[][]{2018Kuncarayakti}, suggesting that other factors more dominant than metallicity could be at play in the production of SESNe. 
The flat distribution over metallicity (from supersolar up to roughly $\sim$ 0.6 $\rm Z_{\odot}$) is also visible in observations of \citet[][]{2012Kelly}. However, we note that in both \citet[][]{2012Kelly} and \citet[][]{2018Kuncarayakti}, the samples come from targeted surveys that preferentially select bright galaxies, which limits the metallicity range covered. The constant rate supports the notion that metallicity may not be a dominant factor in the overall formation of SESNe, consistent with findings from previous studies of SN environments \citep[][]{2010Anderson, 2015Anserson, 2011Leloudas, 2016Galbany, 2024Xi}. 
The SESNe-to-Type II ratio reported by \citet{2023A&A...677A..28P} exhibits substantial fluctuations with metallicity, showing no clear overall trend. The two metallicity calibrations yield different behaviors: using the D16 method, the ratio decreases with metallicity, whereas the M13 method shows an initial decline from 0.39 near solar metallicity to 0.1 at $\sim 0.4~Z_\odot$, followed by a rise at lower metallicities, reaching 0.44. 

We also compare our findings with previous binary population synthesis studies.  \citet[][]{2017aZapartas} find the ratio of SESNe to Type II is not constant but rather decreases with decreasing metallicity. The main reason for this is that they assume that all stripped progenitors lead to a successful explosion, naturally finding higher SESN rate at high metallicities where winds are stronger and produce massive WR stars, that in our simulations do not explode. At the same time, although the rapid population synthesis codes tend to underestimate the re-expansion of partially stripped stars at low metallicities \citep[see][for more details]{2020Laplace}, they find similar SESN rate in $Z\lesssim0.1Z_{\odot}$. 
In their study, the limited information of the progenitor's surface structure does not allow for an accurate distinction between Type Ib and IIb, so they do not report any separate rates from them. 

A similar trend is evident when directly comparing our SESN/II ratio results with those predicted by \texttt{BPASS}. However, both \citet[][]{2022Briel} and the standard \texttt{BPASS} models do not differentiate between Type II subtypes, except for Type IIP events. To estimate the Type IIb rate from the remaining non-IIP Type II events, they adopt a fixed fraction of 0.6541 from \citet[][]{2013Eldridge}. This fraction, however, is poorly constrained and varies across different surveys \citep[e.g.,][]{2011Smith, 2011Li}. Taking this into account, as well as the fact that CC events leading to BH formation are included in \citet[][]{2022Briel}, we find that their predictions generally align with our results at low metallicities (below 0.2 $\rm Z_{\odot}$). At higher metallicities, however, significant discrepancies emerge, with \texttt{BPASS} predicting substantially higher ratios. This may stem from the inclusion of BH-forming SNe and uncertainties in the assumed fraction of Type IIb events.

\subsection{Mass Ejecta of SESNe subtypes with metallicity}

\begin{figure*}
\includegraphics[trim=0 0 0 0, clip=true,width=\textwidth,angle=0]{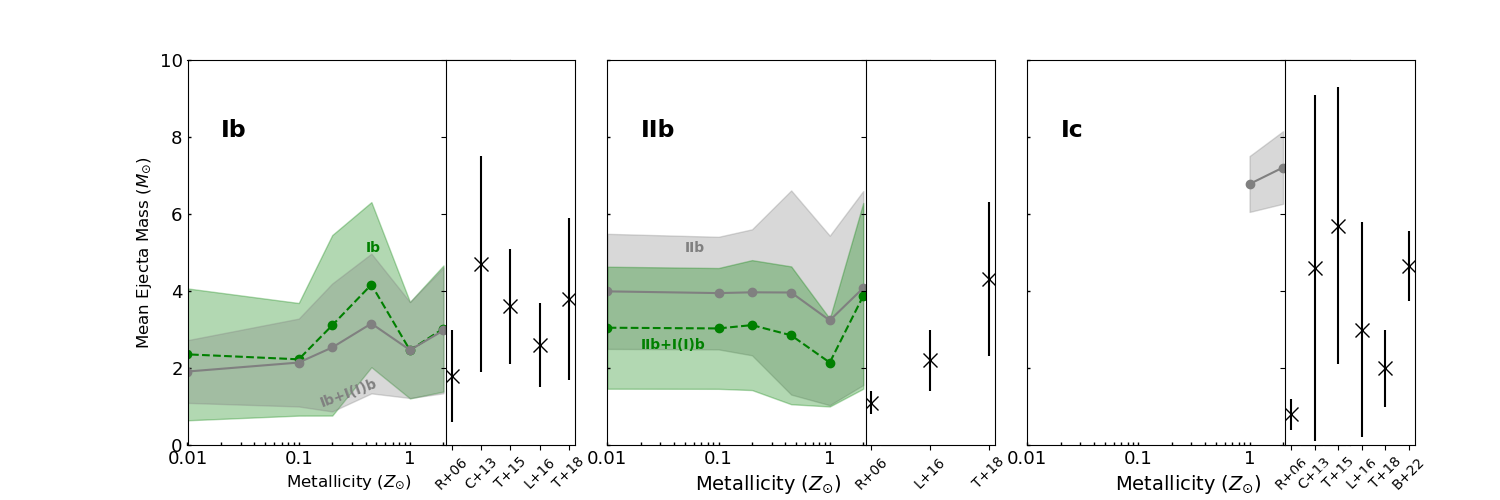} 
\caption {Mean values (solid or dashed lines) and standard deviation (shaded areas) of ejecta masses from our fiducial population for Type Ib, IIb and Ic SNe, at different metallicities. For Types Ib and IIb, we consider two alternative assumptions: (i) grouping the tentative I(I)b class with the IIb SNe (green shading and dashed line in the middle panel), and (ii) treating them as part of the Ib SNe (grey shading and solid line in the left panel).  Comparison with mean ejecta masses among different works in the literature (R+06 = \citet[][]{2006Richard};
 C+13 = \citet[][]{2013Cano}; T+15 = \citet[][]{2015Taddia}; L+16 = \citet[][]{2016Lyman};  T+18 = \citet[][]{2018Taddia}; B+22= \citet[][]{2021Barbarino}).} 
\label{fig:ejecta}
\end{figure*}

Figure \ref{fig:ejecta} shows the mean ejecta masses of different subtypes of SESNe of our fiducial population as a function of metallicity, along with the standard deviations of the distributions. For Type Ib and IIb ejecta masses, we present results under two different assumptions: (i) grouping the tentative I(I)b class together with the IIb SNe (green shading), and (ii) treating them as part of the Ib SNe instead (grey shading).

The average ejecta mass of Type IIb SNe remains roughly constant across metallicities, with $M_{\rm ej} \sim 3.01 \pm 1.70~\rm M_\odot$ when including the tentative I(I)b class, and $M_{\rm ej} \sim 3.86 \pm 1.99~\rm M_\odot$ when I(I)b are excluded, despite the diversity in progenitor origins. At higher metallicities, the standard deviation increases moderately due to the contribution of multiple progenitor channels, including massive single stars, disrupted secondaries, and merger products, in addition to classical binary evolution (see Table \ref{tab:channels}). These alternative channels generally produce more massive ejecta than lower-mass progenitors formed via binary mass transfer at similar metallicities. At lower metallicities (Z < 0.2 $\rm Z_{\odot}$), however, the relative contribution of these channels diminishes because line-driven stellar winds become less efficient, limiting the ability of single stars, disrupted secondaries, and merger remnants to shed their hydrogen-rich envelopes to such an extend to produce type IIb SNe. Below $Z \lesssim 0.2~\rm Z_\odot$, the progenitor population is dominated  primarily by binary systems experiencing case B/C MT, which explains the low ejecta masses and the reduced standard deviation of ejecta masses at low metallicity.

Although metallicity influences binary evolution, its impact on the final ejecta mass is relatively minor due to the compensating effects of MT history. For instance, in a representative binary system with a primary initial mass of 15.7 $\rm M_\odot$, a mass ratio $q_i = 0.7$, and an initial orbital period of 100 days, the post–case B MT hydrogen envelope masses are 0.17, 0.25, and 0.56 $\rm M_\odot$ for metallicities of 0.2, 0.1, and 0.01 $\rm Z_\odot$, respectively, with corresponding helium core masses of 7.25, 7.49, and 8.08 $\rm M_\odot$. While one might expect the ejecta mass to increase at lower $Z$, consistent with trends seen in single-star models \citep[e.g.,][]{2023Aguilera}, subsequent stripping during case C MT reduces the hydrogen envelope mass to 0.12, 0.13, and 0.18 $\rm M_\odot$, and yields helium core masses of 6.69, 6.74, and 6.67 $\rm M_\odot$, respectively. This convergence in both envelope and core masses across metallicity results in broadly similar ejecta masses, despite the differing evolutionary pathways (see also Figure \ref{fig:effect_of_orbital_period} and Table \ref{D2} for more details).

At metallicities above 0.45 $\rm Z_\odot$, Type Ib SNe arise from both low- and high-mass progenitors, predominantly through case B MT (see Section~\ref{sec: metallicity}). In many cases, strong stellar winds then strip away any residual hydrogen envelope after detachment. At $Z \geq \rm Z_{\odot}$, the combined effects of binary MT and stellar winds yield ejecta masses of $M_{\rm ej} \sim 2.75 \pm 1.44~\rm M_\odot$, independent of whether the tentative I(I)b class is grouped with the Ib SNe. As metallicity decreases, the stripping through both winds and MT becomes less efficient, leading to higher ejecta masses that peak at $M_{\rm ej} \sim 3.16 \pm 1.81~\rm M_\odot$ (with I(I)b included) and $M_{\rm ej} \sim 4.17 \pm 2.14~\rm M_\odot$ (with I(I)b excluded) at $0.45~\rm Z_\odot$. Below this metallicity, ejecta masses decline, reflecting shifts in progenitor properties and MT histories. The dominant progenitors become lower-mass stars ($< 12~\rm M_{\odot}$) undergoing multiple MT episodes, such as case B/BB, A/B/BB, B/C/BB or A/B/C/BB (see Section~\ref{sec: metallicity} for more details), which typically produce smaller cores and thus lower ejecta masses.

For Type Ic SNe, we find that the average ejecta masses are substantially higher than those of the other two subtypes, with mean values of $M_{\rm ej} \sim 6.99 \pm 0.83~\rm M_\odot$. However, such events arise only at two metallicities in our models. The larger ejecta masses of Type Ic SNe reflect their origin from the most massive stars in our fiducial population that are still able to undergo CCSNe. This outcome follows from the conservative classification criteria adopted in this study (see subsection \ref{2.1}), under which only the most massive progenitors (with initial masses exceeding $\sim 20~M_\odot$) can successfully shed the bulk of their helium- and nitrogen-rich envelopes, a condition typically satisfied by very massive stars experiencing strong stellar winds in high-metallicity environments.

In Figure \ref{fig:ejecta}, we present the average ejecta masses of different SESN subtypes, as reported in various literature observational studies, though their dependence on metallicity is not yet established. Our models reproduce the observed range of ejecta masses for all subtypes across the full metallicity range. However, our models cannot reproduce the total range of ejecta masses where Type Ic SNe are found. Our models do not produce Type Ic SNe with ejecta masses below approximately 4 $\rm M_{\odot}$, regardless of metallicity.

In summary, our results show that the average ejecta masses of SESN subtypes are either largely insensitive to metallicity (e.g., Type IIb) or exhibit only minor variations (e.g., Type Ib). This trend contrasts with single-star evolutionary predictions, where SESN ejecta masses are expected to increase with metallicity \citep[e.g,][]{2023Aguilera}. We emphasize that the primary driver of this ejecta mass behavior is the occurrence of multiple mass-transfer episodes in binary systems. In particular, the late expansion of partially stripped stars following core helium depletion, which triggers an additional MT episode, plays a key role in efficiently removing most of the H-rich envelope, producing SESNe with low ejecta masses even at very low metallicities (0.1–0.01 $\rm Z_{\odot}$; see Appendix \ref{sec:effect_of_initial_orbital_period}). Therefore, observing these trends in metallicity-dependent studies of SESN ejecta masses can offer valuable insights into their progenitor channels and help evaluate the role and potential dominance of binary systems in producing these events, particularly in metal-poor environments.

\subsection{SESN ages as a function of metallicity}{\label{5.4}}

\begin{figure}
\includegraphics[trim=0 0 0 0, clip=true,width=\columnwidth,angle=0]{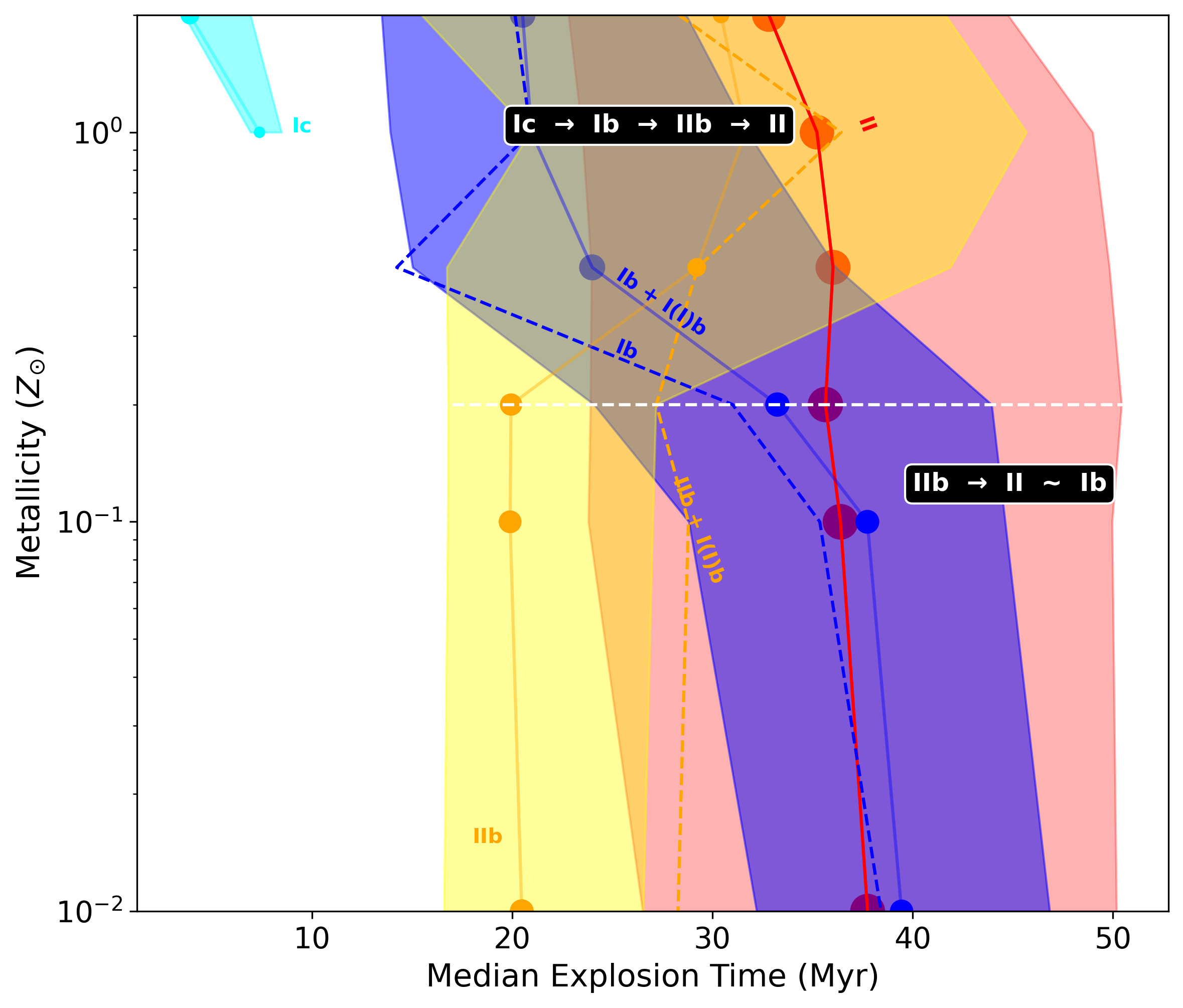} 
\caption {Median explosion times (solid lines) for different SN types as a function of metallicity, with shaded regions indicating the 25th and 75th percentiles. Each color corresponds to a specific SN type: cyan for Type Ic, blue for Type Ib combined with I(I)b, yellow for Type IIb, and red for Type II. Dashed lines indicate the median explosion times for the combined Type IIb + I(I)b and for Type Ib SNe, shown without percentile ranges for improved clarity. The dashed white line marks the metallicity threshold at which the sequence Ic → Ib → IIb → II is altered. Refer to Section \ref{5.4} for further details. } 
\label{fig:explosion_time}
\end{figure}

The evolutionary lifetime of a star is primarily determined by the amount of nuclear fuel available in its core. More massive stars exhaust their fuel much more rapidly than lower-mass stars, resulting in significantly shorter lifespans. In a coeval stellar population formed in an instantaneous burst of star formation from a single, homogeneous molecular cloud, all stars share the same age and metallicity. Consequently, the most massive stars evolve and explode first, followed sequentially by lower-mass stars. Since pre-SN mass loss is strongly correlated with initial mass (with higher-mass stars experiencing stronger winds), the most massive stars more easily shed their hydrogen and helium layers, implying that SESNe progenitors are, on average, more massive than those of Type II SNe. As a result, SESNe should preferentially explode in regions of ongoing star formation, traced by H$\alpha$ emission, embedded within younger stellar populations and denser molecular gas than Type II SNe. It is important to note that this expectation applies primarily to single-star evolution. As demonstrated in the previous section, binary interactions play a crucial role in shaping the evolutionary pathways and progenitor properties of SESNe subtypes across different metallicities. Motivated by these complexities, numerous statistical studies have sought to constrain the masses and ages of different SN progenitors, and in some cases to examine how these properties vary with metallicity, using a variety of age diagnostics. These include analyses of the stellar populations surrounding SNe \citep[e.g.,][]{2014Williams, 2018AWilliams, 2023Sun}, their radial distributions within host galaxies \citep[e.g.,][]{2006James}, molecular gas properties \citep[e.g.,][]{2024Solar}, and associations with H$\alpha$ emission and star-forming regions \citep[e.g.,][]{2017Kangas, 2018Kuncarayakti, 2023A&A...677A..28P}. Despite these efforts, our understanding of the progenitor ages of different SESNe subtypes, and how they vary with metallicity, remains incomplete.

To investigate the impact of binarity and metallicity on SESNe explosion times, Figure \ref{fig:explosion_time} shows the median explosion times for each CCSN subtype from our fiducial population as a function of metallicity, with the 25th and 75th percentiles indicating the spread in the data. At approximately solar metallicity, the shortest explosion times correspond to Type Ic SNe ($\sim$7.38 Myr), followed by Type Ib ($\sim$20.9 Myr, independent of whether the Type I(I)b class is included), Type IIb (31–36 Myr depending on inclusion of Type I(I)b), and finally Type II ($\sim$35.2 Myr). This trend reflects the more rapid evolution and earlier explosion of higher-mass progenitors. Accordingly, the typical progenitor mass is highest for Type Ic SNe and decreases progressively through Types Ib, IIb, and II.

At metallicities below $0.2 \rm Z_\odot$, however, this ordering changes. For instance, at $Z=0.1 \rm Z_\odot$, the shortest explosion times are associated with Type IIb SNe, with median values of 19.9–28.8 Myr. The range depends on the inclusion of Type I(I)b events: when included, the median shifts to higher values, as Type I(I)b generally arise from progenitors with initial masses $\lesssim 12 \rm M_\odot$, while Type IIb originate from more massive stars ($>12 M_\odot$) that at higher metallicities would instead explode as Type Ib (see Section~\ref{sec: metallicity}). Following Type IIb are Type II SNe ($\sim$36 Myr, similar to their solar-metallicity values) and finally Type Ib SNe, with median explosion times of 35–37 Myr depending on the inclusion of Type I(I)b events. This shift indicates that, at low metallicity, Type Ib SNe are predominantly produced by lower-mass stars ($<12 \rm M_\odot$) undergoing multiple mass-transfer episodes (e.g., case B/BB, A/B/BB, or A/B/C/BB; see Section~\ref{sec: metallicity}), which yield smaller cores and longer progenitor lifetimes.

Comparison with environmental statistical studies, which are generally biased toward higher-metallicity environments, supports the sequence of typical progenitor masses predicted by our models at high metallicity \citep[e.g.,][]{2012Anderson, 2014Galbany, 2017Kangas}. However, observing such a sequence does not provide clear evidence about which progenitor channel single or binary is the dominant one, as this sequence can be reproduced also from single stars. Instead, it simply indicates that, on average, SNe Ic arises from more massive progenitors with shorter lifetimes than SNe Ib, IIb, and II, regardless of the binary nature or not.
Additionally, our results show that SNe Ic preferentially occur in higher-metallicity environments, a trend that is also supported by observations. Numerous studies have reported a higher incidence of SNe Ic in metal-rich regions, suggesting that metallicity-dependent stellar winds play a significant role in shaping the progenitor evolution of this particular subclass \citep[e.g.,][]{2018Kuncarayakti, 2012Kelly, 2017Graur, 2010Arcavi, 2016aGalbany}

Conversely, some observational studies suggest trends that diverge from the standard single progenitor mass sequence. For instance, \citet[][]{2012Kelly} reported that the local environments of Type IIb SNe are, on average, bluer than those of SNe Ib, Ic, and II, implying that IIb progenitors may be associated with younger stellar populations and potentially higher initial masses. They also found indications that Type IIb SNe tend to explode in more metal-poor regions compared to SNe Ib and Ic. In a similar vein, \citet[][]{2012Anderson} showed that Type IIb SNe exhibit a strong spatial correlation with star-forming regions within their host galaxies, potentially even stronger than that of SNe Ib, suggesting distinct progenitor properties or evolutionary pathways. Additionally, \citet[][]{2019Xiao} found that SNe II and SNe Ibc originate from progenitors of similar ages across a metallicity range corresponding to 12 + log(O/H) $\sim$ 8.1–8.7, or approximately 0.26 to 1.0 Z$_\odot$. These observed patterns are consistent with our model predictions, which show that both metallicity and binarity can significantly alter the typical progenitor masses of SESNe, something that single star models cannot reproduce.

More recently, \citet[][]{2024Solar}, using measurements of molecular gas at the explosion sites of different SN subtypes, proposed that SNe Ic and SNe II may arise from progenitors of comparable initial mass. This result is not reproduced in our models, likely due to the conservative classification criteria adopted for SNe Ic in this study. It is plausible that some events classified as SNe Ib in our framework would appear as SNe Ic observationally, particularly if nickel mixing is considered, a factor that could help reconcile these findings. Alternatively, this discrepancy might stem from the uncertainties associated with using molecular gas density as a tracer of stellar population age  \citep[see][for further discussion]{2024Solar}.

Additional environmental studies that investigate how this sequence varies with metallicity could offer important insights into the progenitor origins of SESNe and help clarify whether single or binary evolution dominates their formation pathways especially at low metallicity enviroments.

\section{Discussion}\label{sec:6}

\subsection{Winds and metallicity}\label{sec: variations}

A major source of uncertainty in our models lies in the treatment of stellar winds \citep[][]{2014Smith,2017Renzo, 2024Josiek}.
The binary population synthesis models presented in this work adopt the empirical mass-loss rates of \citet[][]{2000Nugis} for partially stripped stars. However, more recent studies, such as \citet[][]{2017Vink}, suggest that low-mass helium stars may exhibit considerably weaker winds. \citet[][]{2019Gilkis} show that adopting such reduced wind strengths, once the surface hydrogen abundance drops below $X_{s}$=0.4, can prevent the complete stripping of the hydrogen-rich envelope, in contrast to the higher rates assumed by \citet[][]{2000Nugis} used in our models. 

Moreover, \citet[][]{2023Gotbrg} report that observed winds from low-mass helium stars in the SMC and LMC (as noted by \citet[][]{2023Drout}) are consistent with or even weaker than the mass-loss rates proposed by \citet[][]{2017Vink}. At solar-like metallicities, such weak winds following case B RLOF can shift the binary evolution pathway from the standard case B to the case B/C RLOF regime. This occurs because the winds may be too weak to completely strip the remaining H-rich envelope after case B RLOF, allowing a thin H layer to survive beyond core helium depletion. This residual envelope might later be removed through an additional mass transfer episode driven by hydrogen shell burning. If this subsequent RLOF fails to fully strip the remaining hydrogen, the system would likely explode as a Type IIb SN rather than a Type Ib, which would be expected under stronger wind conditions. Alternatively, it may remain intact if the envelope mass and binary separation do not favor further interaction, ultimately leading to a Type IIb SN. Consequently, adopting weaker winds for the low-mass partially stripped stars could potentially increase the occurrence of Type IIb SNe especially in high-metallicity environments, where the effects of winds are more pronounced. 

Beyond partially stripped stars, the mass-loss rates of red supergiants (RSGs) can also influence the formation of SESNe. High RSG mass-loss rates have been proposed as a possible explanation for the "red supergiant problem", the observed absence of Type II SN progenitors with initial masses $\ge$ 20 $\rm M_{\odot}$ \citep{2009Smartt}. These strong winds could enable massive stars to shed their H-rich envelopes during the RSG phase, potentially leading to SESNe (if the stars are able to successfully explode) rather than producing typical Type II SNe. However, the physical mechanisms driving RSG winds, and their dependence on metallicity, remain poorly constrained \citep[e.g.,][]{2021Kee,2023Yang, 2024Zapartas, 2024Antonias}. The choice of high, metallicity-independent mass-loss rates in our models, would expect to form more SESNe even in low-metallicity environments. Nevertheless, these strong winds primarily affect more massive stars ($\ge$ 20 $\rm M_{\odot}$), which are less favored due to the initial mass function (IMF) and may be more prone to collapsing directly into BHs rather than exploding as SNe. In addition, some studies showed that RSGs with such winds are rare, suggesting that the stripping mainly occurs through binary evolution or episodic mass loss \citep[e.g.,][]{2022Beasor, 2023Beasor, 2024Decin}. As a result, the overall impact of such winds on the rates of SESN subtypes is likely to be modest. 

Another uncertainty in our stellar wind models is the treatment of the luminous blue variable (LBV)-type winds, which we do not  account for a potential metallicity dependence and use a simple prescription, following \citet[][]{2010Belczynski}. As discussed in \citet[][]{2023Bavera}, LBV-like winds have a limited impact on the evolution of massive stars at high metallicities, where line-driven winds are already strong enough to prevent stars from crossing the Humphreys–Davidson limit. However, at lower metallicities, LBV-like winds may play a more significant role. Due to substantial theoretical and observational uncertainties surrounding LBV winds and their dependence on metallicity \citep[e.g.,][]{2004Owocki, 2006Smith, 2015Meyenet, 2024cheng}, further investigation is needed to better understand the influence of LBV winds on the formation and characteristics of SESNe.

Regardless of the stellar wind assumptions, we expect wind effects to be more pronounced at high metallicities. At low metallicities, wind-driven mass loss is minimal, and mass is primarily lost through RLOF than stellar winds. Therefore, we do not anticipate a substantial impact on our results from wind mass loss in the low-metallicity regime. After exploring six different metallicities, some with metallicity-dependent wind prescriptions, we find that the overall parameter space for SESNe remains largely unchanged across metallicities. This suggests that binary interactions, rather than stellar winds, are the dominant mechanism enabling SESNe especially at low metallicity, where wind effects are limited. However, we emphasize that stellar winds would have a more pronounced impact on the distribution of SESNe subtypes.

\subsection{Robustness of our results and limitations}\label{6.2}

\begin{figure*}
\includegraphics[trim=0 0 0 0, clip=true
, width=\textwidth, angle=0]{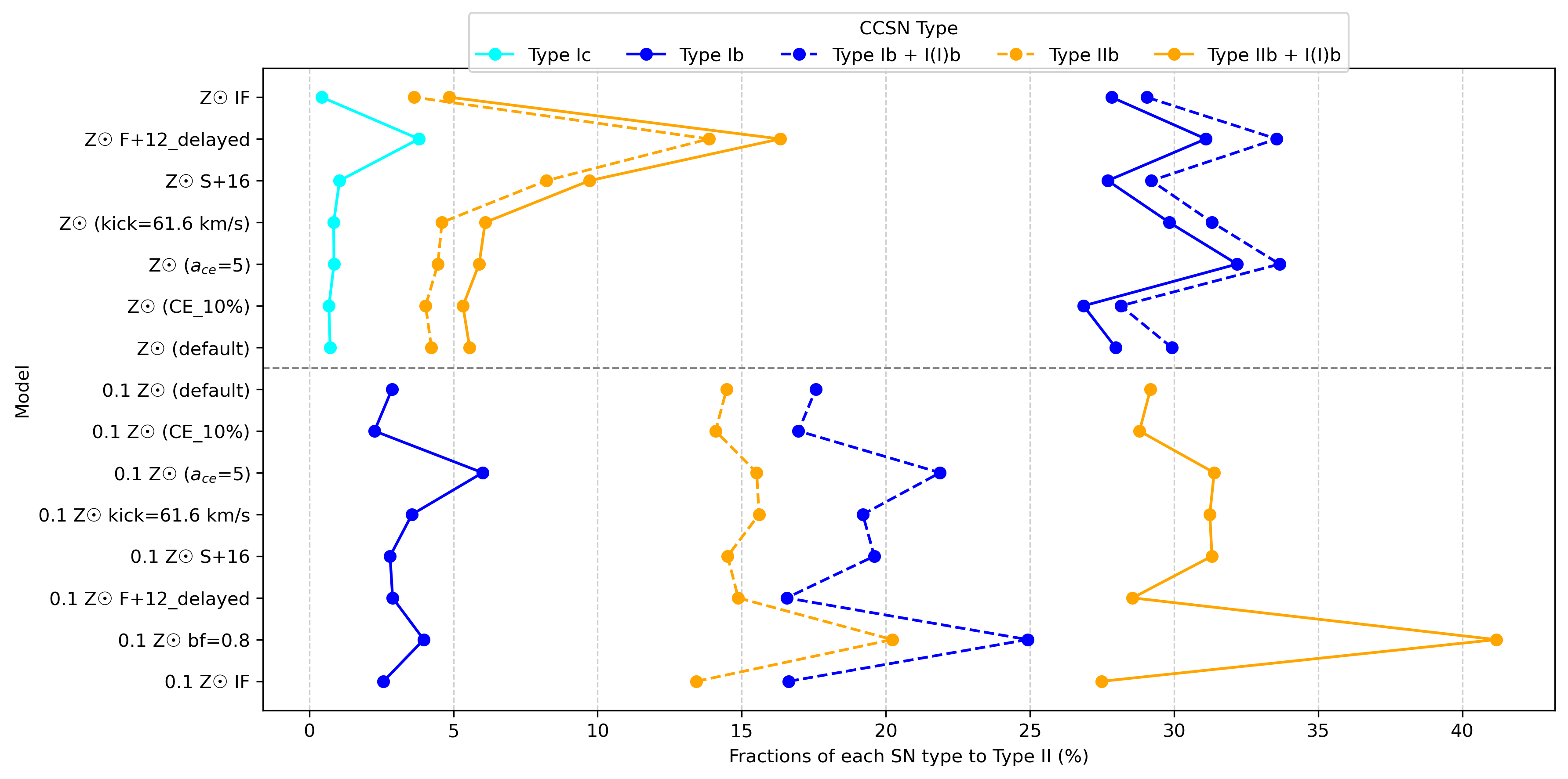} 
\caption {The fractions of SESNe subtypes to Type II for our 15 models, see text in subsection \ref{6.2} for the different model assumptions}
\label{variations}
\end{figure*}

To assess how sensitive our results are to changes in the assumed parameters, we perform a series of simulations that explore different assumptions about physical processes and the handling of initial conditions (see subsection \ref{sec:2}). Specifically, we carry out binary population synthesis calculations for 13 distinct binary population models, varying the binary evolution parameters at metallicities of $\rm Z_{\odot}$ and 0.1 $\rm Z_{\odot}$ (six variations at solar metallicity and seven at subsolar metallicity).

A key factor influencing the production of SESNe, in addition to stellar winds, is CE evolution. In our models, we assume that following CE ejection, the star fully loses its hydrogen-rich envelope, implying that, if it explodes, it will result in a Type Ib or Ic SN. However, simulations indicate that a small amount of hydrogen may survive the CE phase \citep[e.g.,][]{2024Wei}, and whether this residual hydrogen is subsequently lost depends on the strength of post-CE stellar winds. Consequently, our models may overpredict the number of Type Ib/c SNe produced via this channel, especially in low-metallicity environments where weaker winds are less capable of stripping the remaining hydrogen. 

Another limitation of our models concerning CE is that we do not fully account for the effects of "rejuvenation" in secondary stars \citep[e.g.,][]{2023Renzo}. While we allow accretors to spin up rapidly and reach near-critical surface rotation due to mass transfer, capturing some consequences of rejuvenation, we do not follow the full impact of this rapid rotation through the end of our simulations. Such high rotation rates can drive mixing of chemical elements and angular momentum, potentially enlarging the cores of MS accretors and altering their pre-SN core structure. Rotation also influences stellar mass-loss histories, modifies the core-envelope boundary, and significantly reduces the envelope binding energy during the remaining evolution, thereby influencing the CE parameter $\lambda_{\rm CE}$ \citep[e.g.,][]{2025Landri}. However, once one of the two stars collapses, the secondary effectively loses all prior rotational and rejuvenation information. As a result, our models do not include the effects of rejuvenation in secondary stars. Accounting for these effects could potentially increase the number of secondaries that eject their CE, leading to more type Ib/c SNe.

In our standard simulations, a big fraction of Type II SNe result from systems
that merge during a CE phase. Increasing the efficiency parameter for CE ejection $\alpha_{\rm CE}$ enhances
the fraction of systems that avoid coalescence. By increasing the efficiency parameter to five times its default value ($\alpha_{\rm CE}$ = 5), we observe a significant rise in the number of Type Ib SNe, accompanied by a decline in Type II events across both metallicities (see Table \ref{tab:number} where shows the results of our 15 (including the defaults) binary population models, listing
the numbers of each SN Type in our populations). This shift is driven by a higher fraction of binary systems successfully ejecting their envelopes during CE phase, thereby evolving into Type Ib/c SNe. As a result and as displayed in Figure \ref{variations}, the relative fraction of Type Ib to II SNe increases from 27\% to 32\% at solar metallicity, and from 2.86\% to 6\% at subsolar metallicity. Type IIb and Type I(I)b SNe remain largely unaffected by this change, as they do not predominantly form through CE evolution (see subsection \ref{5.1}). Furthermore, since mergers contribute minimally to the formation of Type IIb SNe, the reduced number of merging systems has little effect on their overall occurrence (see Table \ref{tab:channels} for more details). 

We have also explored the impact of the core-envelope boundary which will change the binding energy of the envelope (and thus the outcome of the CE phase) as the deeper envelope layers tend to be the most tightly bound. To assess the impact of the core-envelope boundary definition, we explore two criteria for H-rich stars: the outermost layer where the hydrogen mass fraction falls below 0.3 (default) and 0.1 (model CE$\_$10\% in Figure \ref{variations}). At both metallicities, applying the stricter 0.1 threshold leads to a modest decrease in Type Ib/c SNe and a slight increase in Type II events. This shift arises from a greater number of binary systems failing to eject their envelopes during the CE phase, ultimately leading to mergers. The resulting merged stars, which retain more of their hydrogen-rich envelopes, are more likely to explode as Type II SNe. The ratio of Type Ib to Type II SNe decreases from 27.97\% to 26.86\% at solar metallicity, and from 2.86\% to 2.26\% at subsolar metallicity. Overall, the changes are minimal, suggesting that the Ib/II and IIb/II SN ratios, whether including Type I(I)b with Ib or with IIb, are largely unaffected by variations in the core-envelope boundary assumptions. 

In our models, we employ nearest-neighbour method across all binary populations. To evaluate potential uncertainties from this choice, we also simulate populations at the two metallicities using \texttt{POSYDON’s} initial-final (IF) interpolation method \citep[see subsection 7.2 in][]{2023Fragos}. The comparison shown in Figure \ref{variations} and Table \ref{D2} (model: IF), relative to our default populations at both metallicities, indicates that key qualitative outcomes, such as the number of each SN type and the relative fractions of different SN types compared to Type II, remain consistent. Overall, our quantitative findings demonstrate robustness with respect to the choice of the (interpolation) scheme used within \texttt{POSYDON}.

Stars that lose their outer layers through binary interactions may undergo an ultra-stripped supernova (USSN), typically receiving smaller natal kicks compared to those from iron CCSNe \citep[e.g.,][]{2015Tauris}. Currently, \texttt{POSYDON} does not model USSNe directly. However, we have explored the impact of reduced SN kicks on our results. Reducing the kicks from 250 km/s to 61.6 km/s (see Section \ref{sec:2} for details) produces only a minor effect. This is because the change predominantly affects secondary stars, which contribute less than 20\% of all CCSNe across all metallicities under our default assumptions. Lower kicks increase the likelihood that systems remain bound after the primary explodes. However, these surviving binaries, typically consisting of a NS and a massive companion, often undergo unstable MT if they interact (see subsection \ref{secondaries_se}). Due to their extreme mass ratios, such interactions generally lead to mergers rather than successful SNe. As a result, systems that would have been disrupted and led to a SN at higher kick velocities now instead merge, primarily reducing the number of Type II SNe. Conversely, the number of systems remaining bound with a BH companion increases slightly. However, because most of these systems would have remained bound even at higher kick velocities, the effect on SESN production from secondaries interacting with a BH is modest, resulting in only a slight increase.

We also ran a simulation at 0.1 $\rm Z_{\odot}$ assuming a binary fraction of 0.8, compared to the default value of 0.6. This adjustment is motivated by recent studies suggesting an intrinsically higher close-binary fraction in low-metallicity environments, around 80\%, for early B-type stars (5–15 $\rm M_{\odot}$), where the IMF favors such stars and most CCSNe are expected to originate \citep[][]{2025Villase}. Our results show that SN subtype outcomes are highly sensitive to the assumed binary fraction. The Type Ib/II SN ratio increases from 2.85\% to 3.97\%, or from 18\% to 25\% when including Type I(I)b SNe in the Ib category. Similarly, the fraction of Type IIb SNe rises from 15\% to 20\%, or from 29\% to 41\% if I(I)b events are grouped with IIb SNe. This increase arises because a higher binary fraction leads to a higher fraction of massive stars undergoing binary interactions, which in turn facilitates the formation of SESNe.

Beyond stellar evolution uncertainties, SN explodability in population synthesis relies on simplified methods, either semi-analytic models \citep[e.g.,][]{2012Fryer, 2020Mandel, 2021Schneider} or 1D parameterized simulations \citep[e.g.,][]{2011OConnor, 2016Ertl, 2019Ebinger, 2020Ertl}. These approaches neglect key multi-dimensional effects like neutrino-driven convection and asymmetries \citep[e.g.,][]{2012Janka, 2018Chan, 2020chan, 2018Ott, 2018kuroda, 2019Burrows, 2019Vartanyan, 2020Powell, 2021Vartanyan},
which can impact explosion outcomes. Although such effects could affect SN rates, accounting for them requires computationally expensive simulations, and no prescriptions yet cover a wide range of progenitors. Our study focuses on neutrino-driven explosions; although alternative mechanisms exist \citep[e.g.,][]{1999Mac, 2012Janka, 2016Gilkis, 2019Soker}, they lack general criteria for population-wide application. To test sensitivity to SN prescriptions, we additionally consider the \citet[][]{2016Sukhbold} (calibrated according the “N20" engine) SN prescription (model: S+16) and the delayed model of \citet{2012Fryer} (model: F+12$\_$delayed). We see that our results are most affected by the SN prescription used in our binary population synthesis models at high metallicities. Changing this prescription between F+12$\_$delayed and S+16 impacts significantly the production of Type IIb SNe at high metallicity enviroments,  with the number increasing from 2467 in the default case to 4258 with S+16, and reaching 8410 Type IIb SNe under the F+12$\_$delayed prescription (see Table \ref{tab:number}). This is because, the  S+16 model leads to more optimistic explosion outcomes for SN IIb
at $\rm Z_{\odot}$, consistent with \citet[][]{2021Zapartas}, while in F+12\_delayed model, direct collapse occurs for stars with a $M_{\rm C/O-core}$ $\ge$ 11 $\rm M_{\odot}$, while stars below this threshold are assumed to explode as SNe, resulting in either BH or NS. As a result, the total number of  CCSNe increases significantly, with Type Ic rising from 420 to 2306, Type Ib from 16295 to 18861, and Type IIb from 2467 to 8410, while the number of Type II SNe remains nearly unchanged compared to the default.  This results in a notable increase in the IIb/II SN ratio from approximately 4\% in the default model to 15\% at solar metallicity in the F+12\_delayed model, and from 4\% to 8\% in the S+16 model. However, at lower metallicities, our results are less sensitive to SN prescriptions due to weaker stellar winds and less efficient strpping through mass transfer, which lead to larger C/O core masses, increasing the likelihood of failed SN and direct BH formation. In contrast, at higher metallicities, stronger winds lead to greater mass loss, allowing even very massive stars to develop lower-mass C/O cores (below 11 $\rm M_{\odot}$), which can result in successful CCSNe according to F+12\_delayed.


\subsection{Interacting SNe with circumstellar material and asymmetric explosions}

Since a large fraction of stars in our models fill their Roche lobe and undergo mass transfer up to core carbon depletion (the endpoint of our simulations), particularly at low metallicity, we expect that many of them will still be Roche-lobe filling shortly before CC, or even at the time of explosion. As a result, SN progenitors in binaries may exhibit distorted outer layers (an asymmetric structure) at the pre-SN stage if they are filling, or nearly filling, their Roche lobe. In addition, ongoing mass exchange and interactions between stellar winds and transferred material can generate complex, non-spherical circumstellar environments. These effects are likely to  have implications for the
explosion. Asymmetric explosions have often been proposed as the origin of the polarization signatures observed in many CCSNe \citep[e.g.,][]{2008Wang,2015Mauerhan,2011Dessart,2016Reilly}. Given that such conditions may be common, we highlight the need for detailed modeling of Roche-lobe-filling stars at the time of collapse to test whether envelope distortions can fully explain early-time polarization and to quantify their impact on observable SN properties \citep[][]{2024DuPont}.

Especially at low metallicity, mass transfer in binary systems may continue until core collapse. In our models, if the mass gainer is rapidly spun up, nearly all of the transferred material is expelled from the system through a rotation-enhanced wind. However, key factors such as mass transfer efficiency, ejection velocity, and the composition of the resulting circumstellar material (CSM) at the time of explosion remain uncertain. Since our models are evolved only up to core carbon depletion, we note that some of these systems could potentially be progenitors of certain Type Ibn or Ibc SNe \citep[see e.g.,][]{2007POstarello, 2008Postarello,2022Wu, 2024Ercolino, 2025Ko}. However, a detailed investigation of this possibility lies beyond the scope of this study. In the present work, all such subclasses are grouped under the broader Type Ib and Ic categories.

\section{Conclusions}\label{sec:7}

In this paper, we investigated the evolutionary pathaways that lead to different types of stripped-envelope supernovae (SESNe) as a function of metallicity, using the state-of-the-art binary population synthesis code \texttt{POSYDON} which is based on multi-metallicity extended grids of detailed binary evolution simulations computed by \texttt{MESA}. Our analysis encompasses a wide range of evolutionary channels, including single-star evolution, envelope stripping through stable or unstable (common-envelope) mass transfer from primary and secondary stars (whether the system remains bound or becomes disrupted), and simplified merger products.
By exploring the extended initial binary parameter space and tracing mass-transfer histories through to the final outcomes, we are able to map the full landscape of SESN progenitors. This approach enables predictions of SESN formation rates, subtype distributions, ejecta masses, progenitor ages at explosion, and progenitor properties across different metallicities. Comparison of these predictions with observations provides valuable insight into the formation mechanisms of various types of SESN. Our key findings can be summarized as follows.

1. SESNe originate from both primary and secondary stars across the full range of initial mass ratios, however, secondary stars contribute only $\sim$11\% of the total SESN population across all metallicities. Their contributions to specific SESN subtypes are relatively minor (<16\% for Type Ib, <5\% for Type I(I)b, and <10\% for Type IIb), while they dominate the progenitor population of Type Ic SNe accounting for $\sim$60\% at solar metallicity, primarily via unstable mass transfer onto a BH that survives the common-envelope phase. The majority of SESNe, however, arise from primary stars stripped through stable mass-transfer episodes in binary systems. The common-envelope channel accounts for less than 6\% of SESNe across all metallicities and shows no strong metallicity dependence. This low contribution results from both the lower fraction of interacting stars undergoing unstable mass transfer and the limited number of systems able to survive the CE phase when envelope binding energies are calculated using detailed massive-star models across different evolutionary stages and metallicities, in contrast to previous binary population synthesis studies

2. Our results show that binary interactions, rather than metallicity, primarily determine pre-SN mass loss and the resulting SN types. At low metallicities, most stars that are partially stripped through Roche-lobe overflow  re-expand and undergo additional (post-helium depletion) stable mass-transfer phases, efficiently removing most or all of their remaining hydrogen envelopes. These binary interactions are the main drivers of SESN formation, keeping the overall parameter space largely insensitive to metallicity while also influencing the distribution of SESN subtypes, which remains strongly metallicity-dependent. Consistent with observations, our models reproduce the relatively constant SESN-to-Type II SN ratio across metallicities, yet metallicity significantly affects the relative occurrence of SESN subtypes: the rates of Type Ib/II and Ic/II SNe decline at low metallicity, whereas Type IIb SNe become more frequent.

3. Our results emphasize the critical importance of classification thresholds, particularly the minimum hydrogen envelope mass used to distinguish Type IIb from Type Ib SNe. This criterion has a strong impact on the predicted relative rates of Type IIb and Type Ib SNe, especially at low metallicities. For example, at Z=0.1 \rm $Z_{\odot}$, the Type Ib/II rate can vary from 2.49\% to 17.83\% depending on the adopted threshold. In contrast, at high metallicities, this threshold has little effect, as very few systems fall within the sensitive range and the rates remain largely unchanged. To improve these predictions, further radiative transfer modeling is needed to better determine the observable signatures of progenitors with marginal hydrogen envelopes. 

4. We find that the average ejecta masses of SESN subtypes fall within the observed ranges and show little dependence on metallicity, remaining nearly constant for Type IIb and Ic and varying only slightly for Type Ib, despite the diversity of their progenitors. This behavior is in contrast to single-star predictions. If confirmed observationally, these results provide strong evidence that binary interactions play a dominant role in SESN formation and in shaping their observable properties.

5. By accounting for both binarity and metallicity, our models predict that the age distributions of different SN type progenitors and consequently their progenitor masses vary with metallicity. At solar-like metallicities, Type Ic SNe originate from the most massive progenitors with the shortest lifetimes ($\sim$7.4 Myr), followed by Type Ib ($\sim$20.9 Myr), Type IIb ($\sim$31–36 Myr), and Type II ($\sim$35 Myr). At low metallicities (Z $<0.2 \rm Z_{\odot}$), this sequence changes: Type IIb SNe are linked to the most massive progenitors, while Type Ib SNe arise from the least massive stars capable of core collpase with much longer lifetimes. At intermediate metallicities, Type Ib and IIb SN progenitors have comparable masses and lifetimes. These trends are consistent with statistical environmental studies, which report progenitor mass distributions for SESNe subtypes that cannot be explained by single-star evolution. Further investigation of these metallicity-dependent sequences and trends may provide crucial insights into the role of binarity as the dominant progenitor channel for SESNe.

By implementing a multi-metallicity approach that combines high-accuracy binary modeling with a comprehensive exploration of the binary parameter space, we obtain testable predictions for SESN rates, subtype frequencies, ejecta masses, progenitor ages, evolutionary pathways, and properties across metallicity. Current and upcoming surveys, together with statistical environmental studies of SESNe exploring metallicity dependence, can provide crucial tests to refine our understanding of SESN progenitor properties and the dominant role of binary interactions
in shaping SESNe.

\section*{Acknowledgements}
DS thanks Ylva G\"otberg,  John Antoniadis, Andrea Ercolino, Nobert Langer, Tassos Fragos, Stephen Justham and Selma E. de Mink  for fruitful relevant
discussions.
DS and EZ acknowledge support from the Hellenic Foundation for Research and Innovation (H.F.R.I.) under the “3rd Call for H.F.R.I. Research Projects to support Post-Doctoral Researchers” (Project No: 7933). 
This work is based on observations collected at the European Southern Observatory under ESO programme(s): 096.D-0296 (A), 0103.D-0440 (A), 096.D-0263 (A), 097.B-0165 (A), 097.D-0408 (A), 0104.D-0503 (A), 60.A-9301 (A), 096.D-0786 (A), 097.D-1054 (B), 0101.C-0329 (D), 0100.D-0341 (A), 1100.B-0651 (A), 094.B-0298 (A), 097.B-0640 (A), 0101.D-0748 (A), 095.D-0172 (A), 1100.B-0651 (A), 0100.D-0649 (F), 096.B-0309 (A).
MR acknowledges support from NASA (ATP: 80NSSC24K0932). JJA acknowledges support for Program number
(JWST-AR-04369.001-A) provided through a grant from the STScI
under NASA contract NAS5-03127. MUK was supported by the Swiss
National Science Foundation Professorship grant (PI Fragos, project
number PP00P2 176868)
SG, CL, PMS and ET were supported by
the Gordon and Betty Moore Foundation (PI Kalogera, project numbers GBMF8477 and GBMF12341). 
CPG acknowledges financial support from the Secretary of Universities and Research (Government of Catalonia) and by the Horizon 2020 Research and Innovation Programme of the European Union under the Marie Sk\l{}odowska-Curie and the Beatriu de Pin\'os 2021 BP 00168 programme, from the Spanish Ministerio de Ciencia e Innovaci\'on (MCIN) and the Agencia Estatal de Investigaci\'on (AEI) 10.13039/501100011033 under the PID2023-151307NB-I00 SNNEXT project, from Centro Superior de Investigaciones Cient\'ificas (CSIC) under the PIE project 20215AT016 and the program Unidad de Excelencia Mar\'ia de Maeztu CEX2020-001058-M, and from the Departament de Recerca i Universitats de la Generalitat de Catalunya through the 2021-SGR-01270 grant.

\section*{Data Availability}

As mentioned in Sec.~\ref{sec:2}, the \texttt{POSYDON} v2 data and the options followed in our population predictions are public in Zenodo and GitHub, respectively, for reproducibility and transparency reasons. The output of the population runs will be shared on reasonable request to the corresponding author.



\bibliographystyle{mnras}
\bibliography{WD_bibl} 








\bsp	
\label{lastpage}

\appendix

\newpage
\newpage

\newpage
\section{Parameter Survey of Stripped envelope Supernovae from secondary progenitors at solar metallicity}{\label{a}}

\begin{figure*}
\centering
\includegraphics[trim=0 0 0 0, clip=true, width=\textwidth, angle=0]{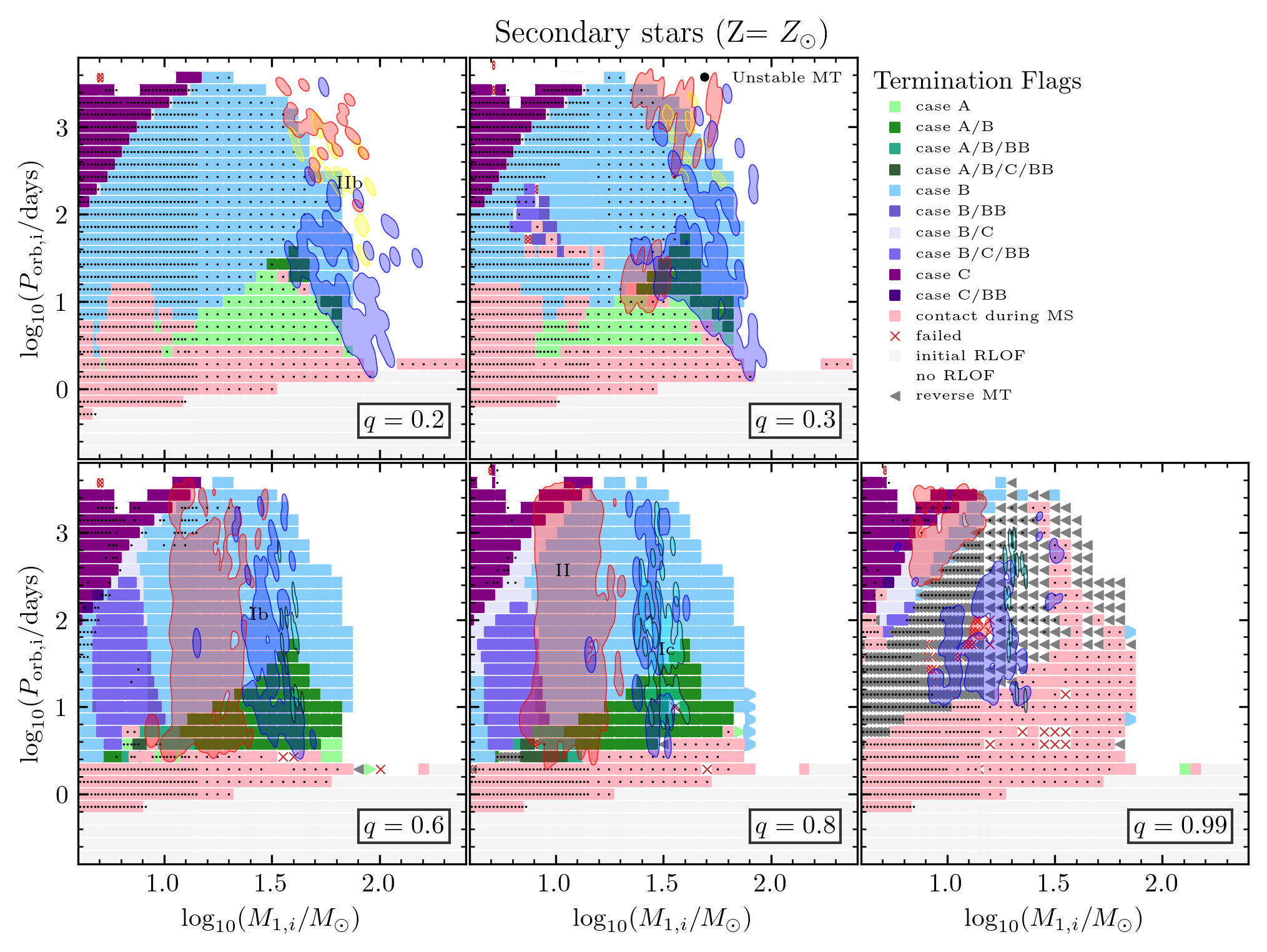} 
\caption{Same as Figure \ref{fig:primary_solar} but we show the parameter space of the predicted final fates of secondaries as a function of the initial orbital period and initial mass of the primary star for fixed initial mass ratios.} 
\label{fig:a1}
\end{figure*}

Figure \ref{fig:a1} shows the predicted final fates of secondary stars as a function of the primary star’s initial mass and orbital period, for fixed mass ratios at solar metallicity. For the parameter space of binaries with 
$M_{1,i}$<25 $\rm M_{\odot}$ and 
$q_{i}\le$0.3, secondaries generally do not produce SNe, either because their initial mass is too low to undergo CCSNe, or because the system merges while both stars are still non-degenerate, effectively removing the secondary from producing a SN. For binaries with  $M_{1,i}$ > 25 $\rm M_{\odot}$, secondaries in systems with initial orbital periods below 100 days typically produce Type Ib SNe, as they remain bound to a BH and lose their outer layers through stable mass transfer (as we have disccussed in Section \ref{secondaries_se}, only about 20\% of the massive binary systems are disrupted by the natal
kick associated with BH formation). At longer periods, Type IIb SNe can arise from secondaries via stable mass transfer onto a BH. Type IIb events are largely confined to lower initial mass ratios, where a BH paired with a low-mass secondary can produce Type IIb or Type I(I)b SNe (see Figure \ref{fig:sec} for the parameter space of Type IIb as a function of the secondary, orbital period and compact object mass). Type II SNe in these mass ratio regimes mostly come from secondaries that are either disrupted or remain bound to a BH without interacting. 

At higher mass ratios $q_{i}>0.3$, Type II SNe from secondaries are associated with the parameter space in which the primary forms a NS (see Figure \ref{fig:primary_solar} for a comparison); because most of secondaries  are disrupted after NS formation (more than $\sim$ 85\% of binaries become unbound due to the natal kick imparted to the
newly formed NS), or for those that remained bound the majority of secondary stars merges with the NS due to the extreme mass ratio, preventing a SN.
For initial primary masses above $\sim$ 20 $\rm M_{\odot}$, secondaries can produce Type Ib or Ic SNe over a wide range of orbital periods, depending on the secondary and compact object masses and the post-kick binary configuration, Type Ib forms primarily through stable mass transfer onto a BH, while Type Ic arises mostly via unstable mass transfer surviving a common-envelope phase (see Table \ref{tab:channels} and Figure \ref{fig:sec} for details). At high mass ratios near unity, there is a region where stable or unstable reverse mass transfer strips the secondary before the primary collapses, in these cases, even if the system is later disrupted, the secondary can still contribute to Type Ib/c SNe depending on its initial mass and subsequent evolution.

\section{Parameter Survey of Stripped envelope Supernovae from primary and secondary progenitors at   0.1 $\rm Z_{\odot}$ }\label{b}

\begin{figure*}
\includegraphics[trim=0 0 0 0, clip=true,width=\textwidth,angle=0]{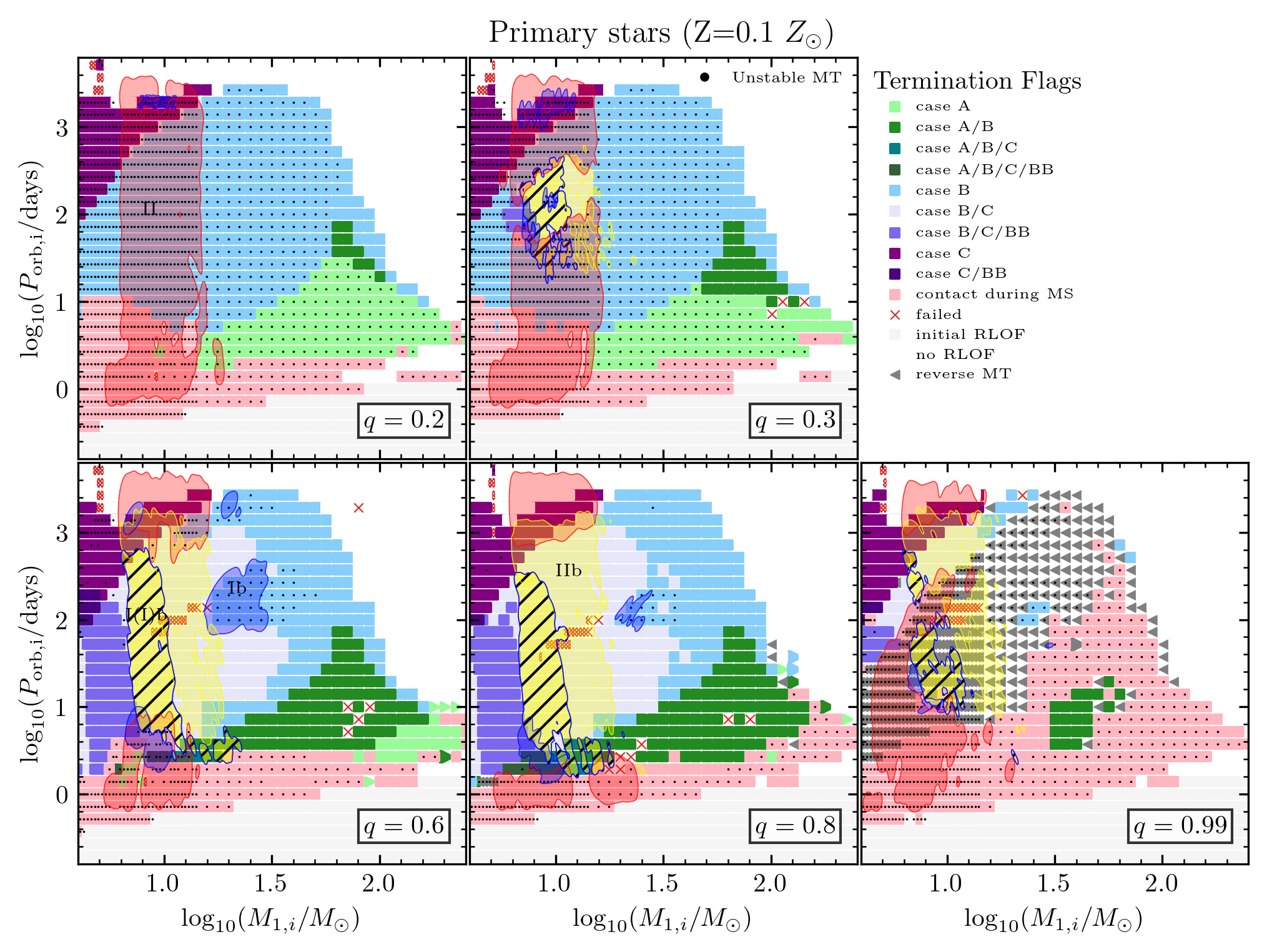} 
\caption {Same as Figure \ref{fig:primary_solar} but for 0.1 $\rm Z_{\odot}$} 
\label{fig:primary_subsolar}
\end{figure*}

Figure \ref{fig:primary_subsolar} shows the predicted final fates of primary stars at 0.1 $\rm Z_{\odot}$. Compared to the solar-metallicity case (Figure \ref{fig:primary_solar}), the parameter space for interacting binaries is substantially larger. At lower metallicities, even the most massive stars tend to expand, leading to stable MT during the MS phase; however, these massive stars generally do not produce CCSNe.

For lower-mass binaries ($M_{1,i} < 30~M_{\odot}$), the MT history of progenitors undergoes notable changes. Most primary stars now produce Type IIb and Type I(I)b SNe through case B/C MT over a wide range of initial masses and orbital periods, although the progenitors of Type IIb are more massive than those of Type I(I)b. The parameter space for Type I(I)b and Type IIb SNe is significantly broader in subsolar metallicity environments, reflecting the reduced efficiency of envelope stripping via both stellar winds and RLOF.

At these low metallicities, Type Ib SNe arise from lower-mass stars ($<12~\rm M_{\odot}$)  primarily through multiple stable MT episodes, such as case B/C/BB or case A/B/C/BB, typically at short initial orbital periods depending on initial mass ratio. Type Ib SNe can also form through successful CE ejection in systems undergoing either unstable case C MT at high initial orbital periods, or in higher-mass primaries (17–30 $\rm M_{\odot}$) experiencing unstable case B MT over a wide range of orbital periods.

Reverse MT in systems with mass ratios close to unity now leads predominantly to Type IIb and Type I(I)b SNe, rather than the Type Ib and Type Ic outcomes observed at higher metallicities. As at solar metallicity, Type II SNe occupy roughly the same overall parameter space; however, the evolutionary channels leading to their production differ. At subsolar metallicity, we find an expanded parameter space for non-interacting binaries that produce Type II SNe compared to the solar case. In contrast, the parameter space for mergers during the contact phase is reduced, since the more compact stellar structures at low metallicity delay the onset of MT to later evolutionary stages, thereby preventing mergers during the MS contact phase.

\begin{figure*}
\includegraphics[trim=0 0 0 0, clip=true,width=\textwidth,angle=0]{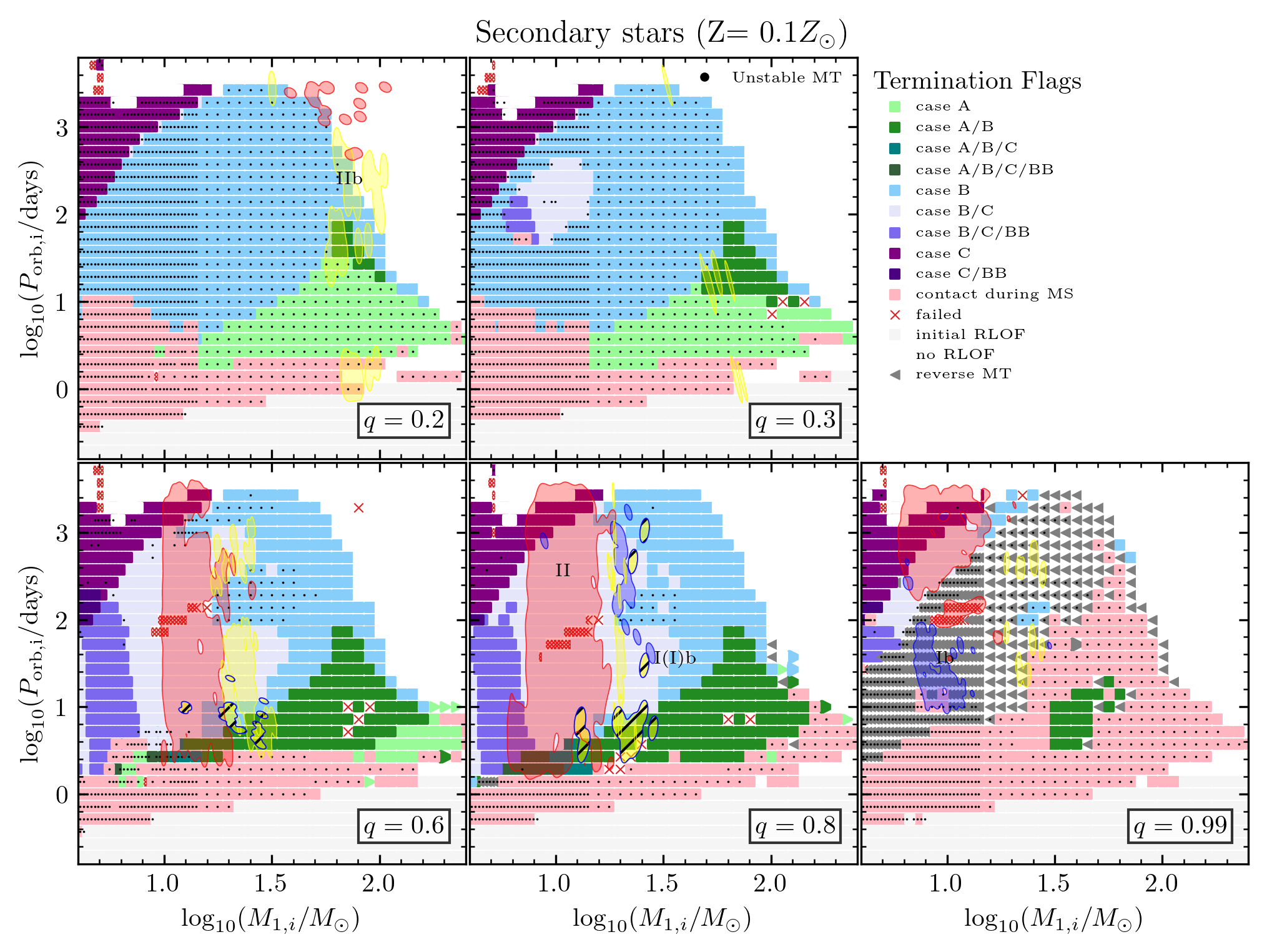} 
\caption {Same as Figure \ref{fig:a1} but for 0.1 $\rm Z_{\odot}$} 
\label{fig:b2}
\end{figure*}

Figure \ref{fig:b2} shows the predicted final fates of secondary stars as a function of the initial orbital period and primary mass at subsolar metallicity. Unlike the solar-metallicity case, Type Ib SNe from secondaries occur mainly at high mass ratios (q>0.7), either through unstable  MT onto a BH that survive the CE, which is confined to systems with  $M_{1,i} > 20~\rm M_{\odot}$, or at even higher mass ratios close to unity, arising from stable reverse MT in systems with $M_{1,i} < 15~\rm M_{\odot}$. For higher primary masses ($M_{1,i} > 15~\rm M_{\odot}$), stable reverse MT from the secondary instead gives rise to Type IIb SNe.

At mass ratios below $\sim$0.9, Type IIb and Type I(I)b SNe are formed primarily through stable MT onto a BH, occupying the parameter space with $\sim M_{1,i} > 12~ \rm M_{\odot}$ across a wide range of initial orbital periods, depending on the mass ratio. As in the solar-metallicity case, most Type II SNe arise in systems where the primary forms a NS and explodes as a SN. This outcome is largely driven by the high disruption rate associated with the natal kicks of newly formed neutron stars, which unbind the binary systems. The surviving secondaries, due to the reduced efficiency of line-driven winds at low metallicity, are unable to shed their outer envelopes and therefore predominantly produce Type II rather than SESNe.

\section{Example of Radius Evolution and Mass Transfer After Core Helium Depletion in Metal-Rich and Metal-Poor Models}\label{sec:radius evolution}

\begin{figure*}
\centering
\includegraphics[trim=0 0 0 0, clip=true, width=\textwidth, angle=0]{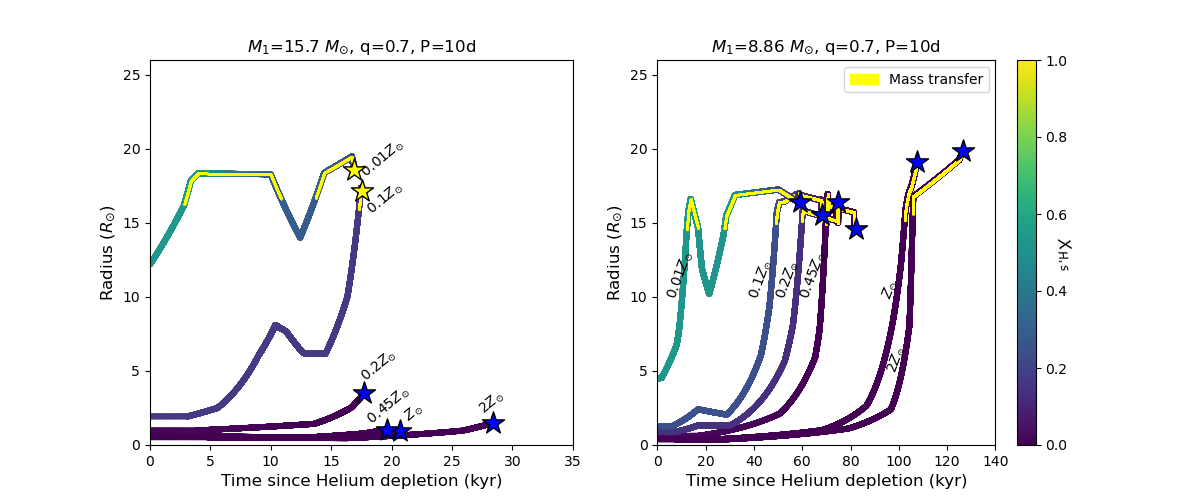} 
\caption{Evolution of the radius as a function of time after core helium depletion for a high (left) and a low mass (right) primary star with initial orbital period of 10 days and initial mass ratio of 0.7 as a function of mettalicity. Colors indicate the remaining surface  hydrogen abundances while yellow thick line indicates stable post-helium depletion mass transfer.} 
\label{fig:radius_evolution}
\end{figure*}

In Section \ref{sec: metallicity}, we discussed that at low metallicities (Z < 0.2 $\rm Z_{\odot}$), the majority of SN progenitors undergo an additional mass transfer episode after core helium depletion. In this section, we first examine the properties and radius evolution since core helium depletion for six representative stellar models (see left panel of Figure \ref{fig:radius_evolution}). These models share the same initial component masses and orbital periods but span six different metallicities. The hydrogen-rich primary star in each model has an initial mass of 15.7 $\rm M_{\odot}$, representing a high-mass SN progenitor within our grid, and is placed in a binary system with an 11 $\rm M_{\odot}$ companion and an initial orbital period of 10 days.

As can be seen in Table \ref{tab:properties} which provides an overview of the key parameters, the amount of the hydrogen-rich layer which retained by the models at the time of core helium depletion (definied at the point where the central helium abundnaces fall bellow $10^{-4}$) increases lowering metallicity. The mass of the hydrogen-rich envelope retained at this point is not only determined by the binary interaction, but can also be affected by wind mass loss. However, at low metallicities where the effect of winds is expected to be limited due to the metallicity dependence of line-driven winds \citep[e.g.,][]{2001Vink, 2017Vink} the binary interaction is mainly responsible for determining the mass of the remaining hydrogen. 

The evolution of the radii is shown in the left panel in Fig. \ref{fig:radius_evolution} as a function of time after core helium depletion for all the primary stars at six different metallicities. For the metal rich models till 0.2 $\rm Z_{\odot}$, we see a monotonic increase of the radius and a non-monotonic increase of the radius for the metal-poor models (0.1 and 0.01 $\rm Z_{\odot}$). 
The difference in the evolution of the radius is linked  to the hyrdrogen mass layer which is retained till core helium depletion \citep[][]{2020Laplace}. Table  \ref{tab:properties}  shows that the high metallicity models have lost all their remaining hydrogen layer by this point due to winds, while the low-metallicity models retain a significant hydrogen layer which can sustain a hydrogen burning shell. The hydrogen shell burning dominates the nuclear luminosity around the
time of core helium exhaustion, and there is a local maximum in the radius evolution (for the 0.1 $\rm Z_{\odot}$ model) or a plateau phase which indicates initiation of mass transfer (for the 0.01 $\rm Z_{\odot}$ model), followed by a contraction and re-expansion phase. The maximum radius each star can achieve at the first expansion phase is determined by the total hydrogen mass retained at the time of core helium depletion. The higher the hydrogen envelope mass, the higher the expansion that they can achieve and are more likely to initiate a mass transfer episode at the first expansion phase. At the peak of the first stellar expansion, hydrogen burning contributes about half of the total nuclear luminosity. At that time the contribution from the helium-burning shell is increasing which leads to the turning point in radial expansion, occurring when the stars have roughly equivalent luminosity contributions from two shell sources \citep[see][for more details]{2020Laplace}. The first expansion in the metal poor-metallicity models is thus associated with hydrogen shell burning and the second is associated with helium shell burning, while the contraction  between the first and second expansion is consistent with the "double mirror effect" \citep[see][for details]{2020Laplace}. Thereafter, both high- and low-metallicity models show radius expansion when helium-shell burning dominates the nuclear luminosity, with the latter ones reaching much larger radii. Thus, low-metallicity models, depending on the retained envelope hydrogen masses, can initiate mass transfer either at the first expansion phase (the case of the 0.01 $Z_{\odot}$ model) or shortly before the moment of the explosion (the case of 0.1 $Z_{\odot}$ model). Here we would like to note that the hydrogen envelope mass that retained at core helium depletion depend on many parameters like metallicity which influence the strength of stellar winds \citep[e.g.,][]{2001Vink} and the evolutionary state of the donor where first  mass transfer occurs and consequently affecting the stripping through mass transfer \citep[e.g.,][]{2020Klencki}, initial orbital period and initial mass ratio \citep[e.g.,][]{2017Yoon, 2024Ercolino}.  

For comparison, in the right panel of Fig. \ref{fig:radius_evolution} we show the radius evolution of a lower-mass SN progenitor (8.87 $\rm M_{\odot}$) placed in a binary system with an initial mass ratio of q = 0.7 and an initial orbital period of 10 days, across six different metallicities. For these low-mass progenitors, late case C/BB mass transfer is more common across all metallicities. This is primarily because stars with final helium core masses below $\sim$ 3 $\rm M_{\odot}$ experience significant expansion during carbon-oxygen core contraction and/or core carbon burning, leading to additional mass transfer episodes, even at the highest metallicities explored (up to 2 $\rm Z_{\odot}$). This expansion occurs regardless of the remaining hydrogen-rich envelope mass at helium depletion. However, at lower metallicities,  expansion is also driven by hydrogen shell burning, which can initiate mass transfer earlier in the evolution, as seen for the 0.01 $\rm Z_{\odot}$ model. This late mass transfer episode for both low and high mass progenitors following core helium depletion is critical, as it impacts the remaining hydrogen envelope mass at the time of core collapse and, consequently, the resulting SN type especially at low metallicity environments (see Appendix \ref{sec:effect_of_initial_orbital_period}).

\section{Impact of Initial Orbital Period on Mass Loss since core Helium Depletion and Its Consequences for Supernova Types at Low Metallicity}\label{sec:effect_of_initial_orbital_period}

\begin{figure*}
\centering
\includegraphics[trim=0 0 0 0, clip=true, width=\columnwidth, angle=0]{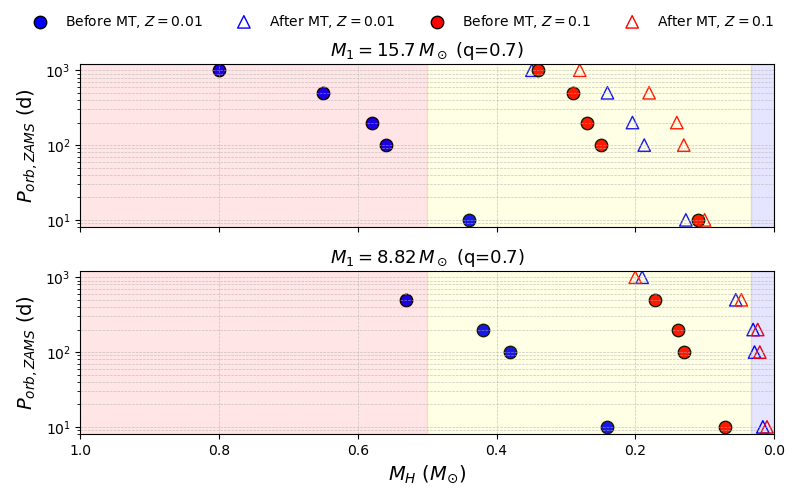} 
\caption{Total hydrogen mass at the time of core helium depletion (filled circles) and at core carbon depletion (open triangles) for low-metallicity models (red color refer to 0.1 and blue color refer 0.01 $\rm Z_{\odot}$) for two primary stars with initial masses of 15.7 and 8.86 $\rm M_{\odot}$, and identical initial mass ratios ($q_{i}$=0.7), plotted as a function of the initial orbital period. Background colors represent the SN type regimes based on the hydrogen envelope mass in the ejecta, Type II (red), IIb (yellow), and Ib (blue). See Table \ref{tab:properties} for more details} 
\label{fig:effect_of_orbital_period}
\end{figure*}

In Figure \ref{fig:effect_of_orbital_period}, we present the total hydrogen envelope mass at two evolutionary stages, core-helium depletion (filled circles) and core-carbon depletion (open triangles), for two primary stars with initial masses of 15.7 and 8.86  $\rm M_{\odot}$, representing high- and low-mass SN progenitors. Both systems have the same initial mass ratio (q=0.7), and the results are shown as a function of the initial orbital period at low metallicities (0.1 and 0.01 $\rm Z_{\odot}$). This comparison illustrates how post–helium-depletion mass-transfer episodes, together with different binary configurations, influence the final hydrogen envelope mass and, consequently, the resulting SN type in low-metallicity environments.

As expected, in both binary systems the mass of the hydrogen  envelope remaining at core-helium depletion increases with wider orbital periods and lower metallicity. However, at both metallicities, the lower-mass primary retains substantially less hydrogen than the more massive primary. (see Section \ref{sec: metallicity} for a detailed discussion). The different background colors in the figure represent the SN type regimes, Type II (red), IIb (yellow), and Ib (blue), based on the hydrogen envelope mass in the ejecta (see classification details in subsection \ref{2.1}).

The post–helium-depletion mass-transfer episode is a key factor in determining the final SN subtype. For the low-mass primary with initial orbital periods below 200 days, this late interaction produces Type Ib SNe instead of Type IIb across both metallicities. At longer orbital periods, the same process shifts the outcome from Type II to Type IIb. For more massive primaries, the influence of post–helium-depletion mass transfer is particularly strong at the lowest metallicity, where it drives nearly all systems that would otherwise yield Type II SNe to instead produce Type IIb explosions.

Interestingly, despite substantial differences in hydrogen mass at core-helium depletion, the final hydrogen mass at core-carbon depletion converges to similar values across models. This suggests that late mass-transfer episodes are highly efficient in stripping the hydrogen envelope and play a decisive role in setting the final SN subtype.

\begin{table*}

\caption{(1) Initial mass of the primary star, (2) initial orbital period, (3) metallicity, (4) initial mass ratio, (5) mass transfer history, (6) total mass at core He depletion, (7) total H mass at core He depletion, (8) total final mass, (9) final radius, (10) final effective temperature in log (K), (11) final luminosity in log ($\rm L_{\odot}$), (12) final C/O core mass, (13) final He core mass, (14) total H mass in the ejecta, (15) SN type.}
\label{tab:properties}
\begin{tabular}{|c|c|c|c|c|c|c|c|c|c|c|c|c|c|c|}

\multicolumn{14}{|c|}{}                  \\ \hline
 $\rm M_{1,i}$   & P   &  Z &  $q_{i}$  &  MT case &  $\rm M_{b} $  & $\rm M_{b,H}$ &  $\rm M_{f}$ & $\rm R_{f}$  &  $\rm  T_{\rm eff} $ & $\rm L_{f}$ & $M_{\rm C/O, f}$& $M_{\rm He, f}$ &$\rm M_{H,ej}$ &  SN type  \\ 
  $(\rm M_{\odot})$  &  (days)  &  ($\rm Z_{\odot}$) &    &   &    ($\rm M_{\odot}$)  &  ($\rm M_{\odot}$) &   ($\rm M_{\odot}$)&  ($\rm R_{\odot}$) &   &  &($\rm M_{\odot}$)& ($\rm M_{\odot}$) &($\rm M_{\odot}$) &

 \\ \hline

8.82  & 10 &  2 &  0.7   &  B/BB  & 2.4  &  2.96e-10
      &  1.84 &  19.85 & 4.24 & 4.51 & 1.47 & 1.84 & 1.57e-39 & Ib (ECSN)\\  
8.82  & 10 &  1 &  0.7   &  B/BB  & 2.6  &  0.00066
      &  2.36 &  19.11 & 4.26 & 4.57 & 1.58 & 2.36 & 5.32e-29 & Ib (ECSN)\\  

8.82  & 10 &  0.45 &  0.7   &  B/C  & 2.85  &  0.017
     &  2.79 &  14.6 & 4.35 & 4.68 & 1.72 & 2.74 & 0.0036 & Ib (CCSN)\\

 8.82  & 10 &  0.2  &  0.7  &  B/C  & 3.09  &   0.04  &    2.97 & 16.38 & 4.33 & 4.72 & 1.85 &  2.9  & 0.0074 & Ib/IIb (CCSN)\\

8.82  & 100 &  0.2  &  0.7  &  B/C  & 3.28  &   0.068  &    3.14 & 107.51 & 3.93 & 4.76 & 1.95 &  3.04  & 0.017 & Ib/IIb (CCSN)\\

8.82  & 200 &  0.2  &  0.7  &  B/C  & 3.38  &   0.091  &    3.19 & 170.68 & 3.83 & 4.77 & 1.978 &  3.09  & 0.021 & Ib/IIb (CCSN)\\

8.82  & 300 &  0.2  &  0.7  &  B/C  & 3.39  &   0.093  &   3.21 & 207.23 & 3.79 & 4.78 & 1.984 &  3.103  & 0.023 & Ib/IIb (CCSN)\\

8.82  & 500 &  0.2  &  0.7  &  B/C  & 3.47  &   0.11  &    3.31 & 319.8 & 3.70 &4.78 & 2.015 &  3.149  & 0.047 & IIb (CCSN)\\

8.82  & 1000 &  0.2  &  0.7  &  C  & 7.89  &   3.19  &   3.77 & 478.04 & 3.63 & 4.83 & 2.17 &  3.39  & 0.191 & IIb (CCSN)\\

8.82  & 10 &  0.1 &  0.7   &  B/C  & 3.27  &  0.07
      &  3.08 &  15.5 & 4.35 & 4.75 & 1.93 & 3.00 & 0.01 & Ib/IIb (CCSN)\\

8.82  & 100 &  0.1  &  0.7  &  B/C  & 3.52  &   0.129  &    3.25 & 107.03 & 3.94 & 4.78 & 2.031 &  3.155  & 0.020 & Ib/IIb (CCSN)\\

8.82  & 200 &  0.1  &  0.7  &  B/C  & 3.56  &   0.138  &    3.28 & 168.19 & 3.84 & 4.79 & 2.043 &  3.173  & 0.023 & Ib/IIb (CCSN)\\

8.82  & 300 &  0.1  &  0.7  &  B/C  & 3.57  &   0.143  &   3.30 & 209.18 & 3.79 & 4.78 & 2.049 &  3.18  & 0.026 & Ib/IIb (CCSN)\\

8.82  & 500 &  0.1  &  0.7  &  B/C  & 3.65  &   0.17  &    3.389 & 317.26 & 3.71 &4.80 & 2.074 &  3.218  & 0.047 & IIb (CCSN)\\

8.82  & 1000 &  0.1  &  0.7  &  C  & 8.09  &   3.31  &    3.84 & 475.99 & 3.63 & 4.84 & 2.22 &  3.439  & 0.20 & IIb (CCSN)\\

8.82  & 10 &  0.01  &  0.7  &  B/C  & 3.75  &   0.24    & 3.22 & 16.4 & 4.35 & 4.79 & 2.04 &  3.11  & 0.016 & Ib/IIb (CCSN) \\

8.82  & 100 &  0.01  &  0.7  &  B/C  & 4.05  &   0.38  &    3.33 & 109.17 & 3.94 & 4.80 & 2.09 &  3.19  & 0.028 & Ib/IIb (CCSN)\\

8.82  & 200 &  0.01  &  0.7  &  B/C  & 4.11  &   0.42  &    3.35 & 171.41 & 3.84 & 4.80 & 2.10 &  3.20  & 0.032 & Ib/IIb (CCSN)\\

8.82  & 300 &  0.01  &  0.7  &  B/C  & 4.14  &   0.44  &   3.36 & 205.29 & 3.80 & 4.81 & 2.11 &  3.204  & 0.035 & IIb (CCSN)\\

8.82  & 500 &  0.01  &  0.7  &  B/C  & 4.29  &   0.536  &    3.43 & 323.50 & 3.71 &4.81& 2.119 &  3.22  & 0.055 & IIb (CCSN)\\

8.82  & 1000 &  0.01  &  0.7  &  C  & 8.73  &   3.78  &    3.85 & 461.36 & 3.64 & 4.84 & 2.26 &  3.43  & 0.19 & IIb (CCSN)

 \\ \hline

 15.7  & 10 &  2 &  0.7   &  A/B  & 4.4  &  1.38e-34
      &  4.34 &  1.48 & 4.92 & 5.01 & 3.13 & 4.34 & 1.37e-36 & Ib (CCSN) \\

 15.7  & 10 &  1 &  0.7   &  B  & 5.30  &  3.41e-38
      &  5.25 &  0.97 & 5.04 & 5.11 & 3.88 & 5.25 & 7.89e-30 & Ib (CCSN) \\ 

15.7  & 10 &  0.45 &  0.7   &  B  & 5.96  &  6.55e-15
      &  5.93 &  1.03 & 5.03 & 5.19 & 4.36 & 5.93 & 3.52e-20 & Ib (CCSN) \\ 

 15.7  & 10 &  0.2 &  0.7   &  B  & 6.56  &  0.035
      &  6.54 &  3.49 & 4.798 & 5.23 & 4.70 & 6.35 & 0.03 & Ib/IIb (CCSN)\\ 

15.7  & 100 &  0.2  &  0.7  &  B/C  & 7.25  &   0.17  &    7.13 & 111.5 & 4.05 & 5.24 & 5.00 &  6.69  & 0.12 & IIb (CCSN)\\

15.7  & 200 &  0.2  &  0.7  &  B/C  & 7.28  &   0.18  &    7.18 & 193.6 & 3.93 & 5.24 & 5.01 &  6.70  & 0.13 & IIb (CCSN)\\

15.7  & 300 &  0.2  &  0.7  &  B/C  & 7.30  &   0.188  &    7.21 & 233.94 & 3.88 & 5.25 & 5.02 &  6.71  & 0.14 & IIb (CCSN)\\

15.7  & 500 &  0.2  &  0.7  &  B/C  & 7.34  &   0.19  &    7.28 & 351.5 & 3.80 & 5.25 & 5.04 &  6.73  & 0.17 & IIb (CCSN)\\

15.7  & 1000 &  0.2  &  0.7  &  B/C  & 7.42  &   0.22  &   7.38 & 432.2 & 3.75 & 5.25 & 5.07 &  6.77  & 0.20 & IIb (CCSN)\\

  15.7  & 10 &  0.1  &  0.7  &  B/C  & 6.95  &   0.11    & 6.93 & 17.13 & 4.45 & 5.24 & 4.88 &  6.52   &0.10 & IIb (CCSN)\\      

15.7  & 100 &  0.1  &  0.7  &  B/C  & 7.49  &   0.25  &    7.21 & 112.37 & 4.04 & 5.25 & 5.05 &  6.74  & 0.13 & IIb (CCSN)\\

15.7  & 200 &  0.1  &  0.7  &  B/C  & 7.53  &   0.27  &    7.26 & 194.01 & 3.93 & 5.25 & 5.07 &  6.75  & 0.149 & IIb (CCSN)\\

15.7  & 300 &  0.1  &  0.7  &  B/C  & 7.56  &   0.28  &    7.28 & 235.36 & 3.88 & 5.25 & 5.08 &  6.76  & 0.158 & IIb (CCSN)\\

15.7  & 500 &  0.1  &  0.7  &  B/C  & 7.60  &   0.297  &    7.35 & 348.83 & 3.80 & 5.25 & 5.09 &  6.79  & 0.183 & IIb (CCSN)\\

15.7  & 1000 &  0.1  &  0.7  &  B/C  & 7.72  &   0.34  &    7.60 & 524.52 & 3.71 & 5.25 & 5.11 &  6.80  & 0.284 & IIb (CCSN)\\

  15.7  & 10 &  0.01  &  0.7  &  B/C  & 7.83  &   0.44    & 7.11 & 18.55 & 4.43 & 5.24 & 5.00 &  6.59   &0.127 & IIb (CCSN)\\ 

  15.7  & 100 &  0.01  &  0.7  &  B/C  & 8.08  &   0.56  &    7.33 & 115.87 & 4.04 & 5.24 & 5.07 &  6.67  & 0.187 & IIb (CCSN)\\

15.7  & 200 &  0.01  &  0.7  &  B/C  & 8.13  &   0.58  &    7.38 & 197.5 & 3.92 & 5.24 & 5.078 &  6.676  & 0.20 & IIb (CCSN)\\

15.7  & 300 &  0.01  &  0.7  &  B/C  & 8.14  &   0.593  &    7.40 & 239.43 & 3.88 & 5.24 & 5.08 &  6.681  & 0.215 & IIb (CCSN)\\

15.7  & 500 &  0.01  &  0.7  &  B/C  & 8.24  &   0.65  &    7.47 & 353.5 & 3.79 & 5.24 & 5.085 &  6.684  & 0.24 & IIb (CCSN)\\

15.7  & 1000 &  0.01  &  0.7  &  B/C  & 8.47  &   0.80  &    7.69 & 532.8 & 3.71 & 5.24 & 5.09 &  6.69  & 0.34 & IIb (CCSN)\\

 \hline
\end{tabular}
\end{table*}

\begin{table*}

\caption{The numbers of each SN type for our 15 population synthesis models (see subsection \ref{6.2}).} \label{D2}
\begin{tabular}{|c|c|c|c|c|c|}

\multicolumn{6}{|c|}{}                  \\ \hline 
 Model  & Ic  &  Ib &  I(I)b  &  IIb & II      \\

 \\ \hline

  0.1 $\rm Z_{\odot}$ IF &- & 1721 & 9346 & 8937 & 66530 \\
  0.1 $\rm Z_{\odot}$ bf=0.8 &- & 2528 & 13327& 12860 &  63602\\
    0.1 $\rm Z_{\odot}$ F+12$\_$delayed &-& 2244 & 10617& 11553& 77626 \\
    0.1 $\rm Z_{\odot}$ S+16 & -& 1702 & 10174 & 8796 & 60579 \\
    0.1 $\rm Z_{\odot}$ kick=61.6 km/s & - & 2228 & 9778 & 9757 & 62531 \\
    0.1 $\rm Z_{\odot}$ ($a_{\rm CE}$=5) &-& 3849& 10166& 9940&  64047\\
    0.1 $\rm Z_{\odot}$ (CE$\_$10\%) &- & 1540& 10032& 9607& 68176 \\
    0.1 $\rm Z_{\odot}$ (default) & - & 1940 & 9956& 9798&  67685 \\
    $\rm Z_{\odot}$ (default) & 420 & 16295&  772 & 2467 & 58250\\
    $\rm Z_{\odot}$ (CE$\_$10\%) & 405 & 16012 & 775 & 2405 & 59614 \\
    $\rm Z_{\odot}$$\rm Z_{\odot}$ ($a_{\rm CE}$=5) & 470 & 17694 & 813 & 2451 & 54975\\
    $\rm Z_{\odot}$ (kick=61.6 km/s) & 462 &  16326&  820 & 2524 & 54730 \\
    $\rm Z_{\odot}$ S+16 & 543 & 14350 & 778 & 4258 &  51790\\
    $\rm Z_{\odot}$ F+12$\_$delayed & 2306 & 18861 & 1495 & 8410 & 60641 \\
    $\rm Z_{\odot}$ IF & 243 & 15813& 692& 2068 & 56810 \\  \hline
\label{tab:number}
\end{tabular}
\end{table*}

\end{document}